\documentclass[apjl]{emulateapj}
\shorttitle{A Counterpart to the Radial Orbit Instability in
 Triaxial Stellar Systems}
\shortauthors{Antonini et al.}
\begin{document}
\def\gap{\;\rlap{\lower 2.5pt
 \hbox{$\sim$}}\raise 1.5pt\hbox{$>$}\;}
\def\lap{\;\rlap{\lower 2.5pt
   \hbox{$\sim$}}\raise 1.5pt\hbox{$<$}\;}

\title{A Counterpart to the Radial-Orbit Instability in
Triaxial Stellar Systems}
\author{Fabio Antonini}
\email{antonini@astro.rit.edu}
\affil{Department of Physics, University of Rome `La Sapienza', P.le A. Moro 5,
I-00185, Rome, Italy, and Dept. of Physics, Rochester Institute of Technology, 85 Lomb Memorial Drive, Rochester, NY 14623, USA }
\author{Roberto Capuzzo-Dolcetta}
\email{roberto.capuzzodolcetta@uniroma1.it}
\affil{Department of Physics, University of Rome `La Sapienza', P.le A. Moro 5,
I-00185, Rome, Italy}
\author{David Merritt}
\email{merritt@astro.rit.edu}
\affil{Department of Physics and Center for Computational Relativity and Gravitation, Rochester Institute of Technology, 85 Lomb Memorial 
Drive, Rochester, NY 14623, USA}

\begin{abstract}
Self-consistent solutions for triaxial mass models are highly non-unique.
In general, some of these solutions might be dynamically unstable,
making them inappropriate as descriptions of steady-state galaxies.
Here we demonstrate for the first time the existence in triaxial
galaxy models of an instability similar to the radial-orbit instability
of spherical models.
The instability manifests itself when the number of box orbits,
with predominantly radially motions, is sufficiently large.
N-body simulations verify that the evolution is due neither to
chaotic orbits nor to departures of the model from self-consistency,
but rather to a collective mode.
The instability transforms the triaxial model into a more prolate,
but still triaxial, configuration.
Stable triaxial models are obtained when the mass contribution of radial
orbits is reduced.
The implications of our results 
for the shapes of dark-matter halos are discussed.
\end{abstract}


\section{Introduction}
Even in a collisionless stellar system, it is possible for density
perturbations to grow, by inducing motions that reinforce the original
overdensity.
Such collective instabilities typically require the unperturbed
motion to be highly correlated, and they have been most
thoroughly studied in thin disks, which are subject to a variety
of instabilities when sufficiently ``cold.''
In elliptical galaxies, where the stellar motions are more nearly
random in direction, density perturbations might be expected to
rapidly attenuate as the stars move along their respective orbits.
However it turns out that the motion in a variety of physically
reasonable models of ``hot'' stellar systems is sufficiently correlated
to induce growing modes \citep[as reviewed by][]{Merritt:1999}.
The two classes of instability that have most thoroughly been studied
in this context are bending instabilities,
which are driven by the centrifugal force of stars moving across
a bend \citep{Toomre:1966}; and the radial-orbit instability (henceforth ROI),
 which
is caused by the tendency of eccentric orbits to clump around
a bar-like distortion 
\citep{Antonov:1973,LB:1979}.
Bending instabilities may be responsible for the lack of elliptical
galaxies more elongated than $\sim 1:3$ 
\citep{PS:1977,MS:1994};
simulated dark-matter halos also appear never to exceed this
degree of elongation, presumably because more flattened halos
``puff up'' due to the instability \citep[e.g.][]{Bett:2007}.
The ROI, on the other hand, can be present
even in precisely spherical galaxies if they contain
an abundance of stars on elongated orbits.
It manifests itself most naturally in collapse simulations,
which produce prolate/triaxial bars if the initial conditions are
sufficiently cold \citep{MA:1985}.
The ROI has also been invoked as a factor
that regulates the density profiles of simulated dark-matter halos
\citep{Bellovary:2008}.
 
Construction of stationary  models of hot stellar systems can
be difficult, and this is one reason why most of the studies cited
above have adopted highly symmetric models: typically spherical,
in the case of the ROI, and axisymmetric
in the case of the bending mode studies.
But there is no reason why such instabilities should be limited
to spherical or axisymmetric models.
Here, we report on a dynamical instability in triaxial models
that closely mimics in its behavior the ROI
of spherical models.
As in the spherical case, the instability leads to a final configuration
that is close to prolate.
Our simulations provide the first concrete evidence that dynamical
instabilities may limit the permitted range of shapes of
triaxial stellar systems, a result that may have 
implications for our understanding of elliptical galaxy
and dark halo dynamics.
 
The self-consistent triaxial models on which our work is based
were described in an earlier paper 
\citep{CLMV07}.
We briefly describe these models in \S 2.
The discretized models are described in \S3,
and the results of $N$-body integrations in \S4 and \S5.
\S6 explores the dependence of the stability properties of
the models on their orbital composition.
\S7 discusses the implications for the dynamics of
elliptical galaxies and dark matter halos.
\S8 sums up.
 
\section{The Self-Consistent Triaxial Models}

The instability was discovered while testing, by $N$-body simulations, 
the equilibrium characteristics of the triaxial galaxy models 
constructed in CLMV07. 
In this section we summarize the way in which the self-consistent
orbital solutions were obtained and in the next section we discuss
the discretized models used in the $N$-body simulations.

CLMV07 constructed three different self-consistent solutions of triaxial, 
cuspy elliptical galaxies embedded in triaxial dark  halos.
The systems differ in terms of the shape of the dark matter halo: 
(i) one halo has the same axis ratios as the luminous matter
(1:0.86:0.7); 
(ii) the second halo has a more prolate shape (1:0.66:0.5); 
(iii) the third halo has a more oblate shape (1:0.93:0.7).  
Our choice was to study the dynamical features of the most interesting 
case of maximal triaxiality (i.e., the model with triaxiality parameter 
$T\equiv (a^2-b^2)/(a^2-c^2)=1/2$), and to study 
the time evolution  of the two self-consistent solutions (MOD1 and MOD1-bis) 
obtained in CLMV07 for this case. 
The main difference between these two solutions was the maximum time 
adopted for the orbital integrations, which was,
in MOD1-bis, longer ($\sim 5$ Hubble times) than in MOD1 
($\sim 2$ Hubble times). 
The models were constructed by means of the  orbital superposition 
method introduced by \citet{Schwarzschild:1979} which relies on an optimization 
technique.
The optimization problem  consisted in minimizing  the discrepancy 
between the model cell masses obtained by integration of the given analytical 
density law $\rho(x,y,z)$  and those given by a linear combination of 
the orbits computed in the potential generated by $\rho$.
In our case, the two quantities to be independently  minimized were
%\begin{mathletters}
\begin{eqnarray} 
\label{op_al_1}
\chi^2_{lm}&=&\frac{1}{N_{cells}} \sum_{j=1}^{N_{cells}}\left( M_{j;lm}- \sum_{k=1}^{n_{orb}}C_{k;lm}B_{k,j;lm}\right)^2 , \\
\label{op_al_2}
\chi^2_{dm}&=&\frac{1}{N_{cells}} \sum_{j=1}^{N_{cells}}\left(M_{j;dm}- \sum_{k=1}^{n_{orb}}C_{k;dm}B_{k,j;dm}\right)^2 ,
\end{eqnarray}
%\end{mathletters}
where $B_{k;j;lm(dm)}$ is the fraction of time that the $k$th orbit spends in the $j$th cell of
the luminous-matter grid (dark-matter grid); $M_{j;lm(dm)}$ is the mass which the model places in the $j$th
cell of the luminous-matter grid (dark-matter grid).
$C_{k;lm}$ and $C_{k;dm}$    represent the total  mass, respectively, of 
the luminous and dark matter component assigned to the $k$th orbit ($ 1\le k \le n_{orb}$).
The basic constraints were  $C_{k;lm}\ge 0$ and  $C_{k;dm} \ge 0$,
i.e., non-negative orbital weights.
The departure  from self-consistency was measured in CLMV07 by
\begin{equation} \label{self-con}
\delta=\frac {\sqrt{\chi ^2}}{\overline M}~~,
\end{equation} 
where  $\overline M$ is the average mass  contained in the grid cells and $\chi ^2$ are the quantities
defined above; thus $ \delta$ represents the fractional rms deviation in the cell masses.

The mass model considered  in CLMV07 for the luminous component was 
a triaxial generalization of Dehnen's (1993) spherical model
with a ``weak''  inner cusp, $\rho\sim r^{-1}$. 
The luminous mass density  was
%\begin{mathletters}
\begin{eqnarray}
\displaystyle \rho_{l}(m) &=& \frac{M_l}{2 \pi
a_{l}b_{l}c_{l}} \frac{1}{m (1+m)^3} \\
\displaystyle
m^2&=&\frac{x^{2}}{a_{l}^{2}}+\frac{y^{2}}{b_{l}^{2}}+\frac{z^{2}}{c_{l}^{2}}
, \qquad 0 < c_{l}< b_{l}< a_{l}
\end{eqnarray}
\label{rhol} 
%\end{mathletters}
and $M_l$ the total luminous mass. 
For the dark component the adopted mass density  was
%\begin{mathletters}
\begin{eqnarray} 
\rho_{dm}(m')&=&\frac{\rho_{dm,0}}{(1+m')(1+{m'}^2)} \\
{m'}^2&=&\frac{x^2}{a_{dm}^2}+\frac{y^2}{b_{dm}^2}+\frac{z^2}{a_{dm}^2}
\end{eqnarray}
\label{den_dm} 
%\end{mathletters}
and $\rho_{dm,0}$ the central dark matter density \citep{Burkert:1995}.
Therefore the dark component  has a low-density core. 

In the present work we adopt the same units used in CLMV07: 
$G=a_l=M_l=1$. Consequently, the time unit is:
%\begin{mathletters}
\begin{eqnarray}\label{unitatempo}
\displaystyle \left[T\right]=G^{-1/2}{a_l}^{3/2}{M_l}^{-1/2}\\
\displaystyle 
= 1.49 \times 10^6 {\rm yr}\Big(\frac{M_l}{10^{11}M_{\odot}}\Big)^{-1/2}\Big(\frac{a_l}{1 {\rm kpc}}\Big)^{3/2}{}.
\end{eqnarray}
%\end{mathletters}
 The, derived, velocity and energy units are $V_u=\sqrt{GM_l/a_l}$ 
and $E_u=(G{M_l}^2/a_l)$, respectively.
In this units the half mass crossing time of the system is:
%\begin{mathletters}
\begin{eqnarray}\label{tempo_cr}
 \displaystyle t_{cross}=\left(\frac{G\left(M_l+M_d\right)}{{r_h}^3}\right)^{-1/2}
= 17.78\left[T\right]
%=26.49 \times 10^6 {\rm yr}%\Big(\frac{M_l}{10^{11}M_{\odot}}\Big)^{-1/2}\Big(\frac{a_l}{1 {\rm kpc}}\Big)^{3/2}{},
\end{eqnarray}
%\end{mathletters}
where $r_h$ is the radius containing half of the model mass, considering
the dark matter halo truncated at $r=80$; $M_d$ is the total mass of the dark matter component.
In the following, this time will be considered the reference time scale.
%\begin{mathletters}
%\begin{eqnarray}\label{unitatempo}
% \displaystyle T=\left(\frac{G\left(M_l+M_d\right)}{{r_h}^3}\right)^{-1/2}=\left(\frac{G\left(11.00M_l\right)}{\left(15.11a_l\right)^3}\right)^{-1/2}
% \\
%\displaystyle
%= 26.49 \times 10^6 {\rm yr}\Big(\frac{M_l}{10^{11}M_{\odot}}\Big)^{-1/2}\Big(\frac{a_l}{1 {\rm kpc}}\Big)^{3/2}{},
%\end{eqnarray}
%\end{mathletters}
%where $r_h$ is the radius containing half of the model mass, considering
%the dark matter halo truncated at $80a_l$; $M_d$ is the total mass of the dark matter component.
%The resulting energy unit is  $E={a_l}^2M_l/T^2$.
%\begin{mathletters}
%\begin{eqnarray}\label{unitaen}
% \displaystyle E=\left({a_l}^2M_l/T^2\right)=\left(\frac{M_l}{10^{11}M_{\odot}}\right)\left(\frac{a_l}{1 {\rm kpc}}\right)^{2}
%\left(\frac{T}{26.49 \times10^6 yr}\right)^{-2}
%\end{eqnarray}
%\end{mathletters}
%$M_{l}=10^{11}M\odot$,
%$a_{l}=1$kpc and $t_{\rm cross}=26.5$ Myrs, 

\section{Discretized Models and their Properties}

In this section we explain the methods we used to discretize
the self-consistent models described above and how
we computed their properties for the $N$-body simulations.
We also present some kinematical features of the models
that are relevant to their stability properties.

\subsection{Discretization}

The initial conditions for the $N$-body integrations were set 
by populating the generic $k$th orbit with a number of particles  
proportional to $C_k$  and randomly choosing positions and 
velocities from the recorded data of the orbital integrations.
In more detail:

\begin{itemize}
\item[1]The values $C_{k}$  for both the dark matter and  the luminous matter are read from  model data and the 
following quantities are evaluated:
%\begin{mathletters}
\begin{eqnarray}
N_{k;lm}&=&C_{k;lm}/m_{lm}~~, \\
N_{k,dm}&=&C_{k;dm}/m_{dm}~~, \\
N_{k}&=&N_{k;lm}+N_{k;dm}~~,
\end{eqnarray}
%\end{mathletters}
where $N_{k}$ is the total number of particles that populate the 
$k$th orbit while $N_{k;lm}$, $N_{k;dm}$ are the 
number of stars and dark matter particles, respectively;
$m_{dm}$ and $m_{lm}$ are free parameters that specify
the mass of individual star and dark matter particles.

\item[2]$N_{k,dm}$ particles with masses $m_{dm}$,
and $N_{k,lm}$ particles with masses $m_{lm}$,
are selected with positions and velocities drawn 
uniformly and randomly (with respect to time) from
the stored positions and velocities of the orbit integration.

\item[3]If $N_{k}$ is greater than the number of available data 
in  ``initial data'',   
further positions and velocities are assigned using a cubic spline 
interpolant.

\end{itemize}
\noindent

These steps are repeated for all orbits of the self-consistent model.
In this way the mass distributions of the models are  adequately reproduced 
as long as the various orbits are populated with a sufficiently large number 
of particles.

\begin{figure}
\includegraphics[angle=270,width=.5\textwidth]{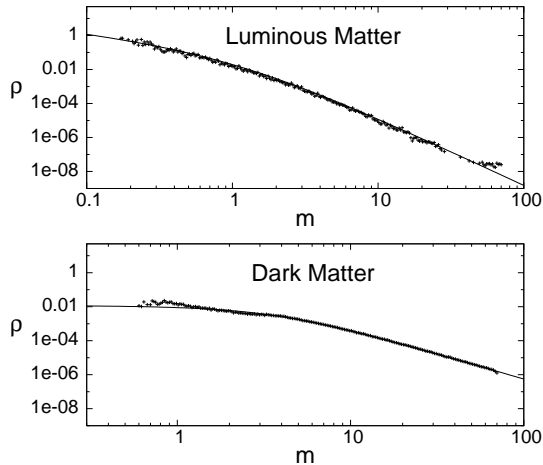}
\caption{Density profiles of the discretized model 
MOD1-bis plotted versus elliptical radius $m$;
the  curves are the analytic input profiles. }
\label{fig1}
\end{figure}

\begin{figure*}
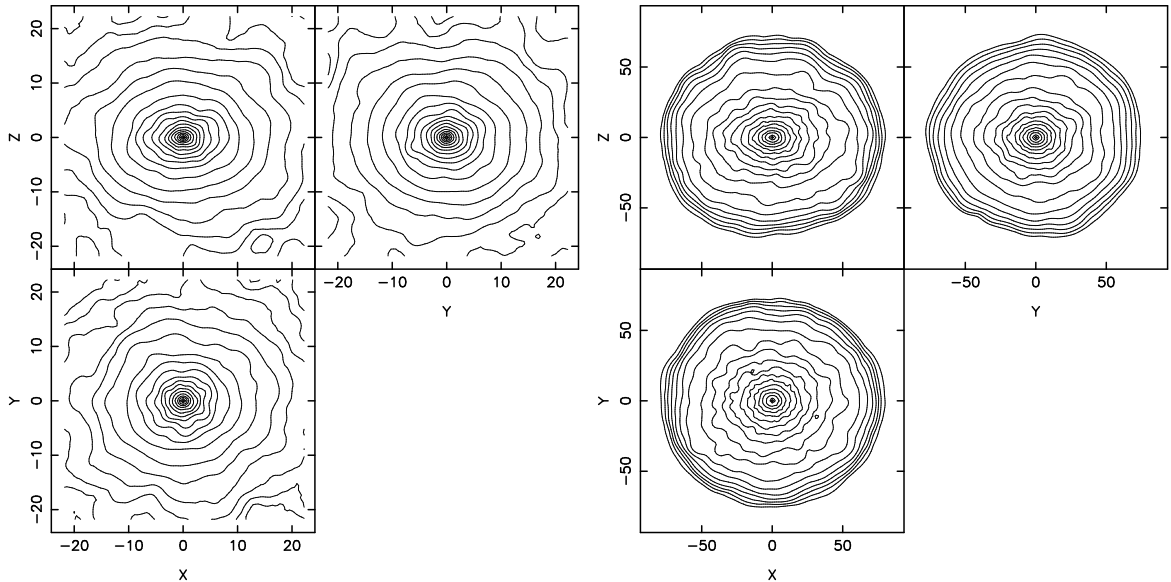

\begin{center}
$\begin{array}{cc}
                \includegraphics[angle=0,width=3in]{fig2a.ps}  & \includegraphics[angle=0,width=3.in]{fig2b.ps} 
                    \end{array}$
\caption{Contours of the projected density for the $N$-body
initial conditions illustrated in Figure~\ref{fig1} for the
luminous matter (left) and the dark matter (right).}
\label{fig2}
\end{center}
\end{figure*}

The total number of objects  depends on  $m_{dm}$ and $m_{lm}$ as 
%\begin{mathletters}
\begin{eqnarray}
N_{dm}&=&{1\over m_{dm}}\sum_{k=1}^{n_{dorb}}C_{k;dm}, \\
N_{lm}&=&{1\over m_{lm}}\sum_{k=1}^{n_{lorb}}C_{k;lm}. 
\end{eqnarray}
%\end{mathletters}
We chose to assign the same mass to each particle of the same type
(luminous or dark matter), hence different numbers of particles are 
spread on orbits with different values of $C_{k}$. 
However, different values were chosen for the mass associated
with dark and luminous components. 
We  took $m_l=5\times 10^{-5}$ and $m_{dm}=7\times 10^{-5}$ which gave,
 for instance in the case of MOD1-bis, 
a total of $166194$ particles, of which $19709$ 
were ``stars''  and $146485$ ``dark matter''; 
the  mean number of particles per orbit was  $36$. 
With this resolution the theoretical mass 
profiles were well reproduced as shown in Figures~\ref{fig1} 
and~\ref{fig2} for model MOD1-bis.

\subsection{Determination of shapes}

The evolution of the shape of the $N$-body systems was studied by 
computing the axis ratios at different distances from the center,
 and also by constructing isodensity contours.

The symmetry axes  were determined from the inertia tensor as
\begin{equation} \label{rapp_ax}
\zeta=\sqrt{T_{11}/T_{max}}~,~~\eta=\sqrt{T_{22}/T_{max}}~,~~\theta=\sqrt{T_{33}/T_{max}}
\end{equation}
where $T_{ii}$ are the principal moments of the inertia tensor
and $T_{max}={\rm max}\{T_{11},T_{22},T_{33}\}$. 
Referring to a coordinate system in which the inertia tensor is diagonal, 
the assumption $\zeta=1$ (i.e., $T_{11}=T_{max}$)  implies 
\begin{equation}
\eta=\sqrt\frac{{ \sum m_i y_i^2}}{\sum m_i x_i^2~},~~ \theta=\sqrt\frac{{ \sum m_i z_i^2}}{\sum m_i x_i^2~}~~.
\end{equation}
The axis ratios of the models were computed through the  standard 
procedure described by \citet{Katz:1991} (similar methods
are described by \citet{DC:1991} and \citet{PM:2004}).  
To evaluate the system shape within a sphere of radius $d$, 
the following iterative method was used:
\begin{itemize}
\item[1.] The inertia tensor defined by particles within a sphere of 
radius $d$ is calculated.
\item[2.] The axis ratios are determined from equation~(\ref{rapp_ax}).
\item[3.] New axis ratios are computed considering only particles 
enclosed in the ellipsoidal volume having the axis ratios determined 
in step 2. 
Therefore, a particle \emph{i} is included in the  summations if  
$q_i<d$, where
\begin{equation} 
q_i^2=\Big{(}\frac{x_i}{\zeta}\Big{)}^2+\Big{(}\frac{y_i}{\eta}\Big{)}^2+\Big{(}\frac{z_i}{\theta}\Big{)}^2.
\end{equation}
\end{itemize}
\noindent
These three steps are  iterated until the axis ratios converge.
Finally, we defined $a>b>c$  assuming $c/a={\rm min}\{\zeta,\eta,\theta\}$
and $b/a$ the intermediate value between ($\zeta,\eta,\theta$).
Evaluation  of the axis ratios of the discretized models verified
the accuracy of the technique:
for the solution MOD1-bis we found $b/a=0.86$, $c/a=0.69$  
for the luminous matter at $r=12$ and $b/a=0.86$, $c/a=0.72$ 
for the dark matter at $r=25$, compared with the  given values
$b/a=0.86$, $c/a=0.70$ of the analytical density law.

\begin{figure}
\begin{center}
$\begin{array}{cccc}
                \includegraphics[angle=270,width=3.5in]{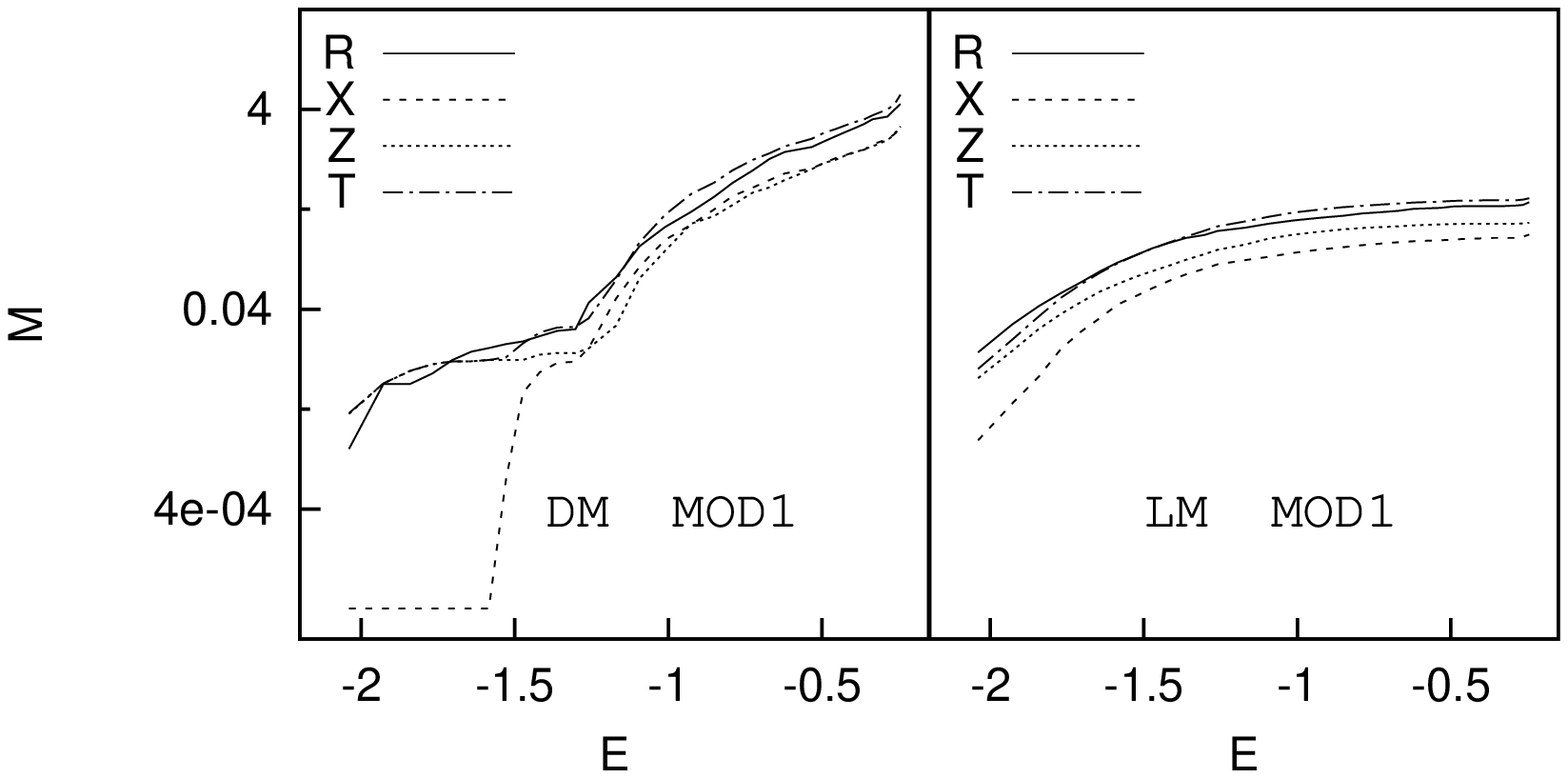}  \\
                \includegraphics[angle=270,width=3.5in]{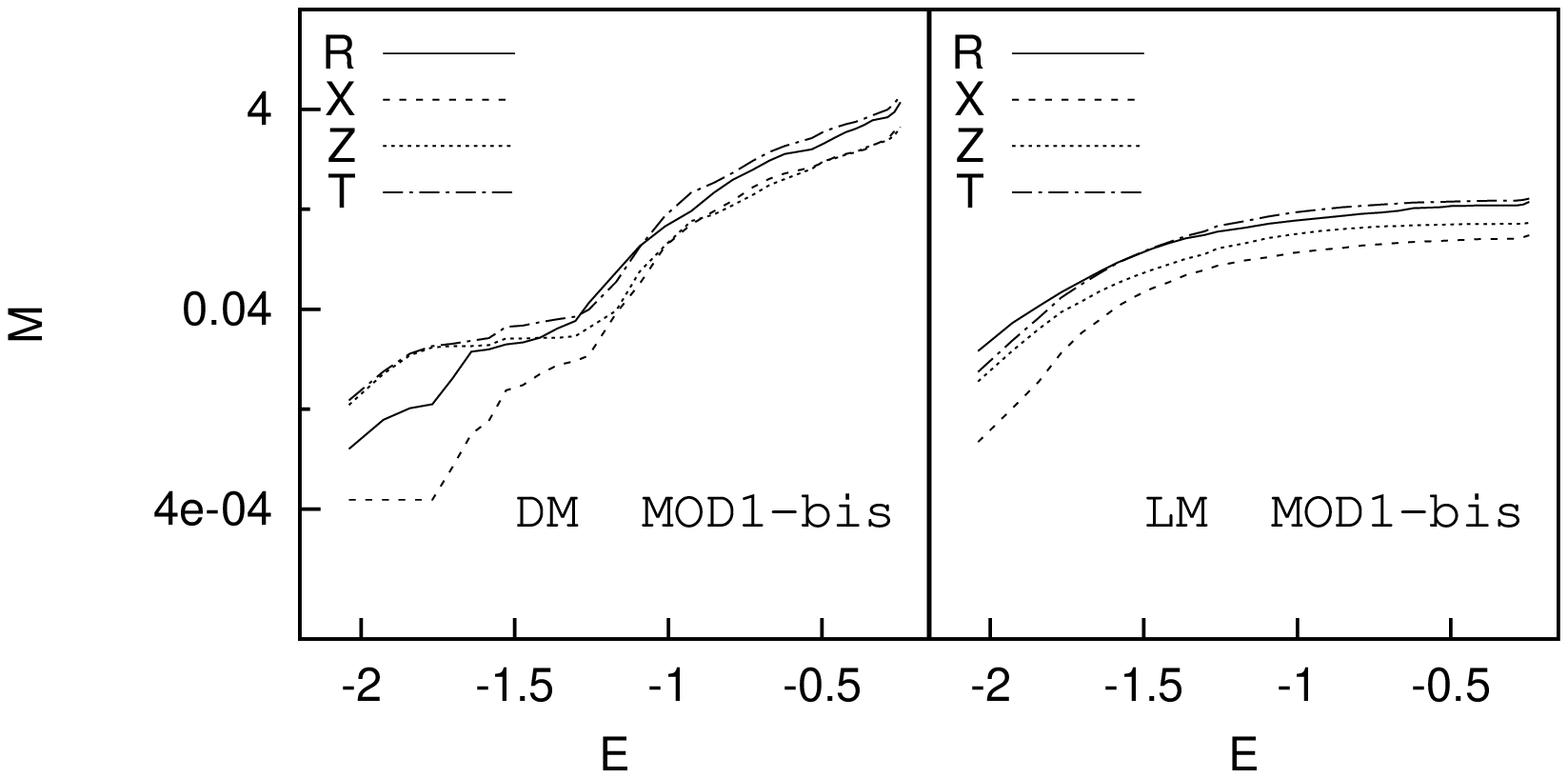}  
    \end{array}$
\caption{
Cumulative (by mass) energy distributions of the various orbital 
families for the dark matter (DM) and the luminous component (LM),
in the self-consistent solutions $MOD1$ (top) and $MOD1-bis$ 
(bottom).
The symbols ``R'', ``X'', ``Z''  and ``T'' denote the mass contributed 
by semi-radial, $X$-tube, $Z$-tube and tube orbits, respectively.}\label{fig3}
\end{center}
\end{figure}

\subsection{Specification of streaming motions}

$X(Z)$-tube orbits are here defined as those orbits having a 
non-vanishing $x(z)$ component of the time-averaged angular momentum;  
all other, non-tube orbits (either box or chaotic) are defined as 
\emph{semi-radial orbits}.
Figure~\ref{fig3} shows the cumulative energy distributions of the 
various orbital families in the discretized models.
There are significant contributions from both tube and semi-radial 
orbits in both the luminous and dark components. 

A choice must be made concerning the sense of rotation of particles
placed initially on tube orbits \citep{Schwarzschild:1979,Merritt:1980}.
The time-averaged density of an orbit is invariant to a change in sign 
of the initial velocity;
maximum rotation (i.e. streaming) is obtained if all particles on 
each tube orbit have  the same sense of rotation, 
while zero mean motion is achieved by populating the tube orbits
equally in both directions.
To investigate the effects of non-zero streaming, 
we constructed two discretizations of each self-consistent solution 
having the two extreme cases of maximum and minimum net streaming. 
In Table \ref{table1}  some parameters of four $N$-body systems, 
sampling MOD1 and MOD1-bis, are given. 

\begin{table}
\caption{Features of $N$-body models.\label{table1}} 
\begin{tabular}{lllll}
%\multicolumn{6}{c}{\textbf{TABLE 1}} \\ 
\tableline\tableline
$System$                & $Solution$     & $L$       &  $N_{lm}$  &  $N_{dm}$     \\\tableline 
$HL$                & $MOD1$         & $23.71448$     &  $19684$ &  $144886$   \\
$HL_{bis}$                & $MOD1-bis$     & $23.57945$    &  $19709$ &  $146485$   \\
$LL$                & $MOD1$         & $0.401937$   &  $19684$ &  $144886$   \\
$LL_{bis}$                & $MOD1-bis$     & $0.337428$  &  $19709$   &  $146485$       \\
\tableline
\end{tabular}
\tablecomments{The acronyms $HL$ and $LL$
stand for ``high $L$'' and \\``low $L$'' where $L$
is the absolute value of the angular momentum.}
\end{table}

\subsection{Model kinematics}
The kinematical features of the discretized models were analyzed by computing 
the first and the second moments of the stellar distribution function
on a spatial grid \citep[e.g.][]{Merritt:1980}.

We used different, Cartesian grids for the two different 
matter components.
In the case of the \emph{dark matter grid}  the cells were cubes with 
sides of   length $6$  and all cells had the same size.
Because of the high density concentration of the luminous component,
we used grid cells with a range of sizes for the \emph{luminous matter grid}.
This grid consisted of a set of cubic cells with sizes ranging 
from $0.5$ near the center to $2$ at greater distances. 

A total of  nine quantities were averaged for both matter
components in each cell:  
the mean velocity $\langle V_i\rangle$ and the six independent components of 
the tensor $\langle V_iV_j\rangle$. 
In this way the velocity dispersion tensor,
\begin{displaymath}
\sigma_{ij}^2 \equiv \langle V_iV_j\rangle - 
\langle V_i\rangle\langle V_j\rangle,
\end{displaymath}
could be evaluated in each cell.
Then, $\sigma_{ij}^2$ was diagonalized obtaining the three 
``principal''  dispersions, and the three direction cosines 
giving the directions of the eigenvectors.

\begin{figure*}
\begin{center}
$\begin{array}{cc}
 \includegraphics[angle=270,width=3.1in]{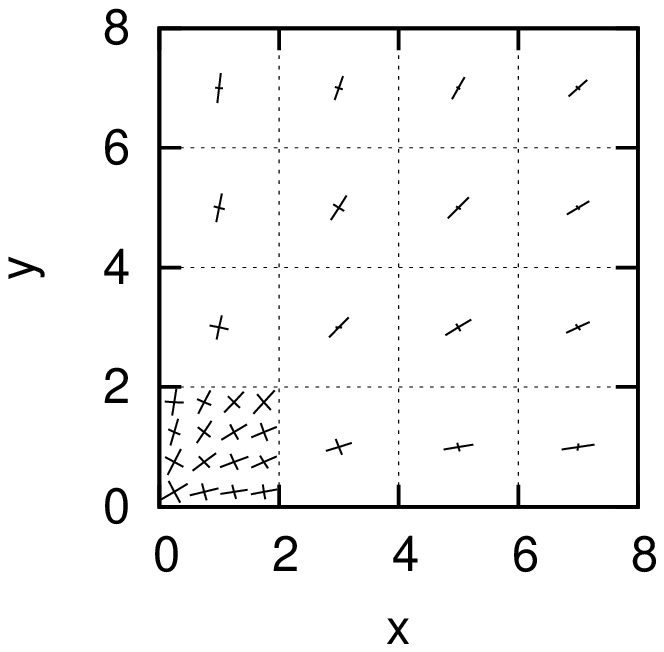} &\includegraphics[angle=270,width=3.1in]{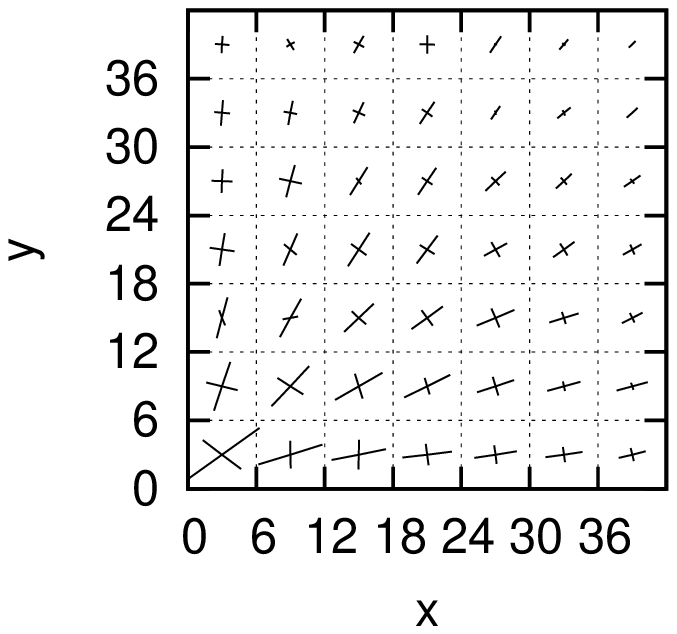} 
\end{array}$
\caption{\footnotesize{Velocity dispersions of  $MOD1-bis$ in the case of luminous matter (left) and for
the dark matter (right). 
The components of the velocity dispersions were stored in each cell of a grid divided as described in the text.
The length of the arms of each cross is proportional to the corresponding principal dispersion.}}\label{fig4}
\end{center}
\end {figure*}

Figure~\ref{fig4} shows the velocity anisotropy in the $x-y$ plane 
for MOD1-bis in the case of high angular momentum.
The  length of each cross arm is proportional to the principal value 
of $\sigma_{ij}^2$. 
Some important features are: 
(1) a high degree of anisotropy in both components at all radii; 
(2) a nearly constant (as a function of radius) radial velocity 
dispersion of the luminous matter.  On the other side, the radial 
velocity dispersion of the dark matter decreases strongly with radius.
 
We evaluated the ``anisotropy parameter'' $2T_r/T_t$ where 
$T_r=\langle v_{r}^2/2\rangle$ and $T_t=\langle v_{t}^2/2\rangle$
(in full isotropy, $2T_r/T_t=1$).
We stress that the interpretation of  this parameter, 
which is straightforward in spherical geometry,   is more
complicated in the triaxial case.
Nevertheless, they give some indication of the average degree
of velocity anisotropy.
All of our discretized models yielded about the same values for the 
anisotropy parameters:
 $\left(2T_r/T_t\right)_{dm}\sim 2$  and  $\left(2T_r/T_t\right)_{lm}\sim 1.4$. 
The high degree of ``anisotropy'' in the dark component -- too
large to be accounted for simply in terms of the triaxial geometry --
suggests a strong bias toward radial motions in the dark matter halo,
as indeed can be seen in Figure~\ref{fig4}.

\section{Fixed-Potential Integrations}

Following \citet{SM:1982}, a useful technique for checking whether
a discrete, $N$-body representation of an equilibrium model was
correctly constructed
is to integrate the discretized initial conditions in the {\it fixed}
potential of the analytic model and observe whether there is
any change in the spatial distribution of the $N$ particles.
This procedure is also a powerful way to constrain the nature
of any evolution that is observed in the full $N$-body integrations.
Actually, if the shape of a model changes with time,
this could be due either to chaotic evolution of individual
orbits (mixing), or to collective modes (dynamical instability).
But the former mechanism will be active even when the potential
is fixed, whereas a collective mode requires an evolving potential.

Accordingly, we integrated the orbits of the discretized models
in the analytic potential corresponding to the smooth mass
distributions (equations~\ref{rhol},\ref{den_dm}).
Each particle was advanced, up to 30 crossing times, 
using a $7/8$ order 
Runge-Kutta algorithm described by \citet{Fehlberg:1968} 
with a variable time step, in order to keep the relative error per step 
in energy less than a specified value ($10^{-8}$).
Since each orbit is independent, this operation is easily parallelized.
The simulation required  $\sim 5$ CPU hours total for  $166000$ particles.
Registration of the particle positions and velocities were made at fixed 
intervals of time, the same for all particles. 
The duration of $30t_{cross}$ is longer than the time over which
the instability manifests itself in the full $N$-body simulations
(see below).

Figure~\ref{fig5} shows that 
no significant evolution of the axis ratios is observed in the fixed 
potential integrations. Also the contours of the projected density for 
both systems remain approximately unchanged until the end of the 
integrations. Actually, the relative variations of the axis ratios with 
respect to their initial values are within $4\%$ for the luminous component and 
 $1\%$ for the dark matter. 
Given  the  unavoidable noise in the computation of the axis ratios,  
such variations are irrelevant;  
the larger variations in the luminous matter are probably a 
consequence of the higher noise due the lower number  of particles. 
These  results allow us  to  conclude that the initial conditions
were correctly generated, and also that any  strong global shape 
deformations in the full $N$-body simulations must be a manifestation 

of a dynamical instability, and not chaotic mixing of individual orbits.
\begin{figure}
\begin{center}
$\begin{array}{c}
                \includegraphics[angle=270,width=2.6in]{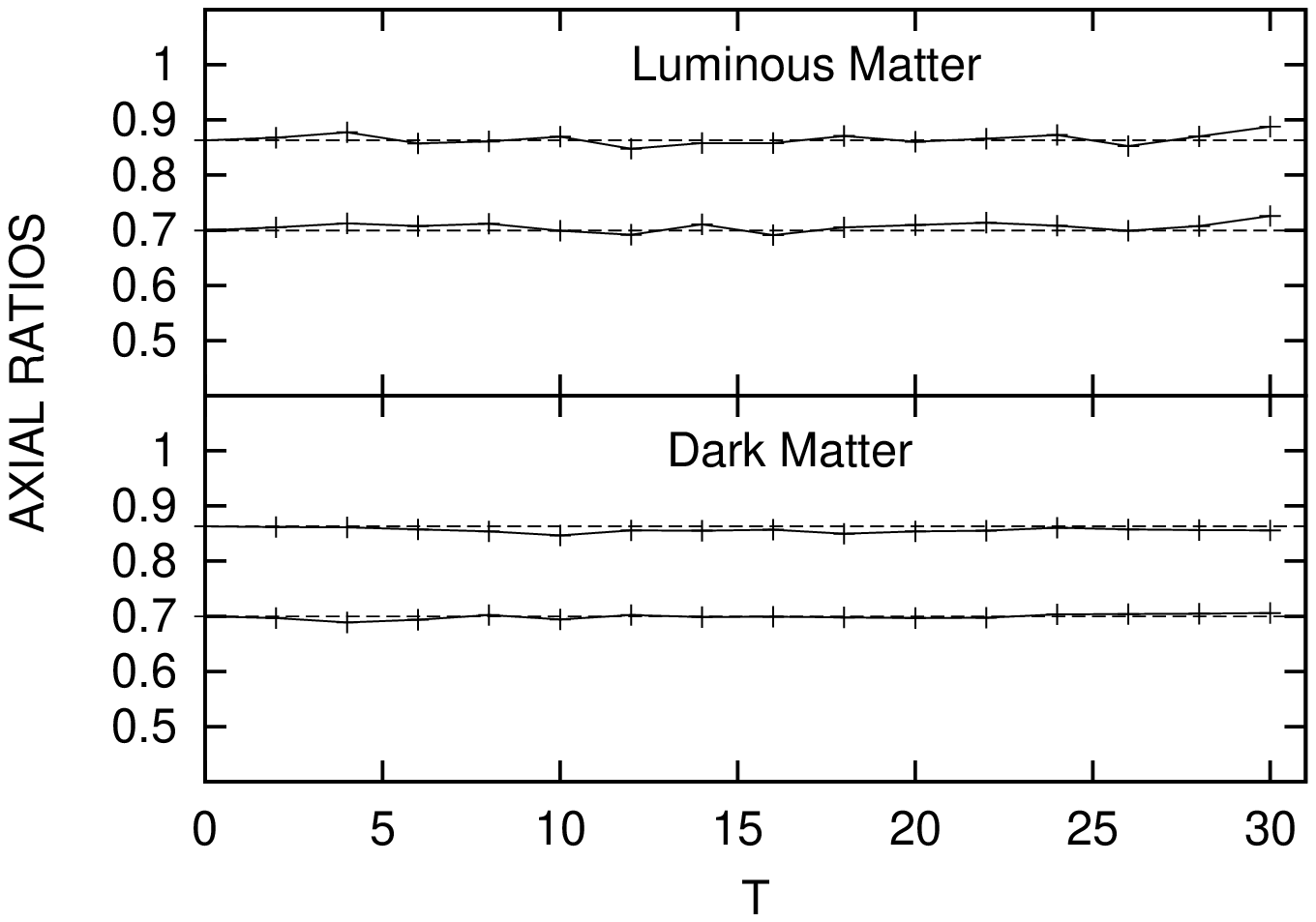} \\
                \includegraphics[angle=270,width=2.6in]{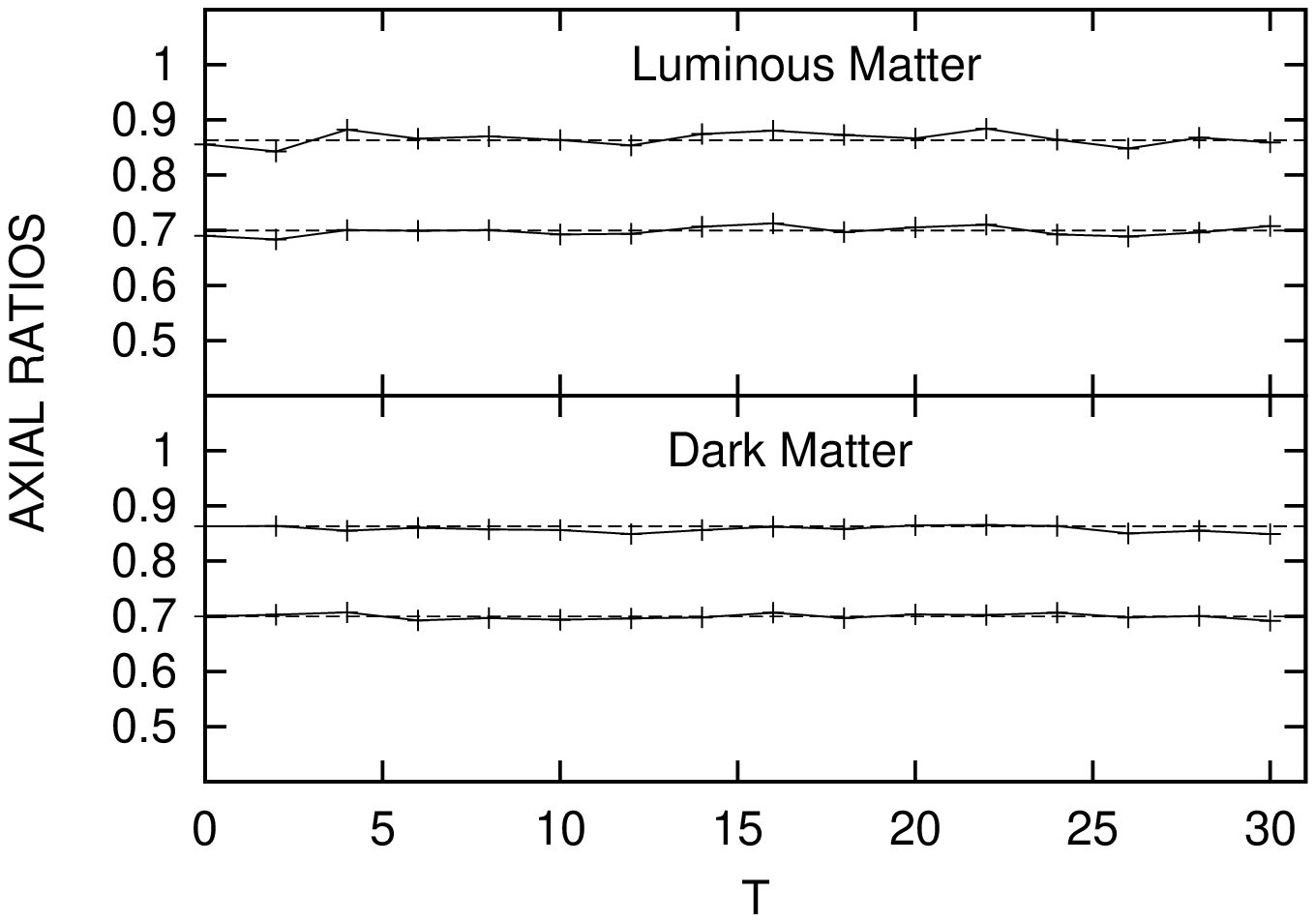} 
    \end{array}$
\caption{
Upper panels: evolution of the axis ratios for the solution MOD1-bis in 
the fixed-potential integrations.
Lower panels: evolution of the axis ratios for the solution MOD1.
The axis ratios are evaluated at $r = 8$ for the luminous matter  
and at $r = 25$ for the dark component. The times are scaled to the internal 
crossing time.}\label{fig5}
\end{center}
\end{figure}

\begin{figure*}
\begin{center}
$\begin{array}{cc}
                \includegraphics[angle=270,width=3.in]{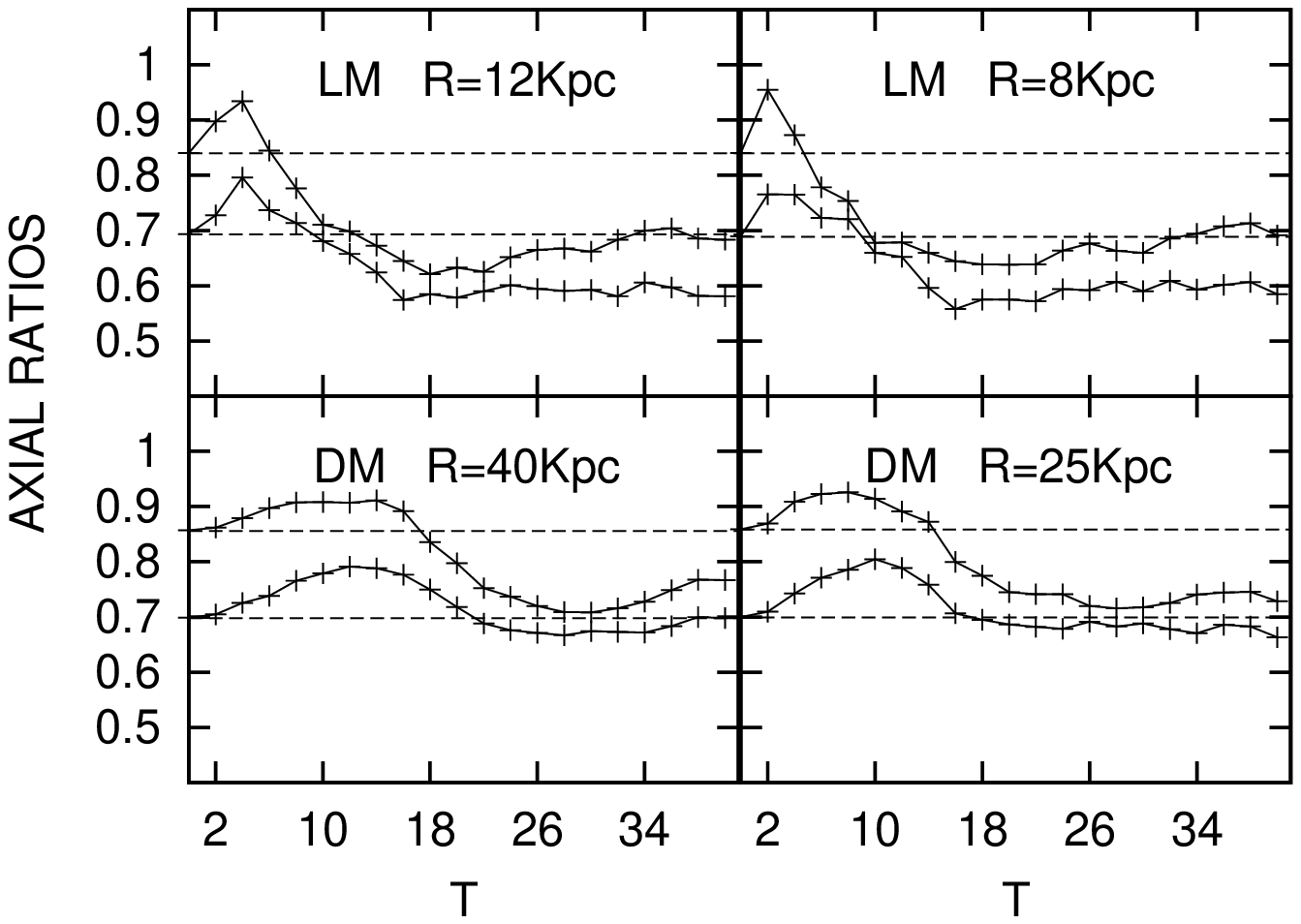}  & \includegraphics[angle=270,width=3.in]{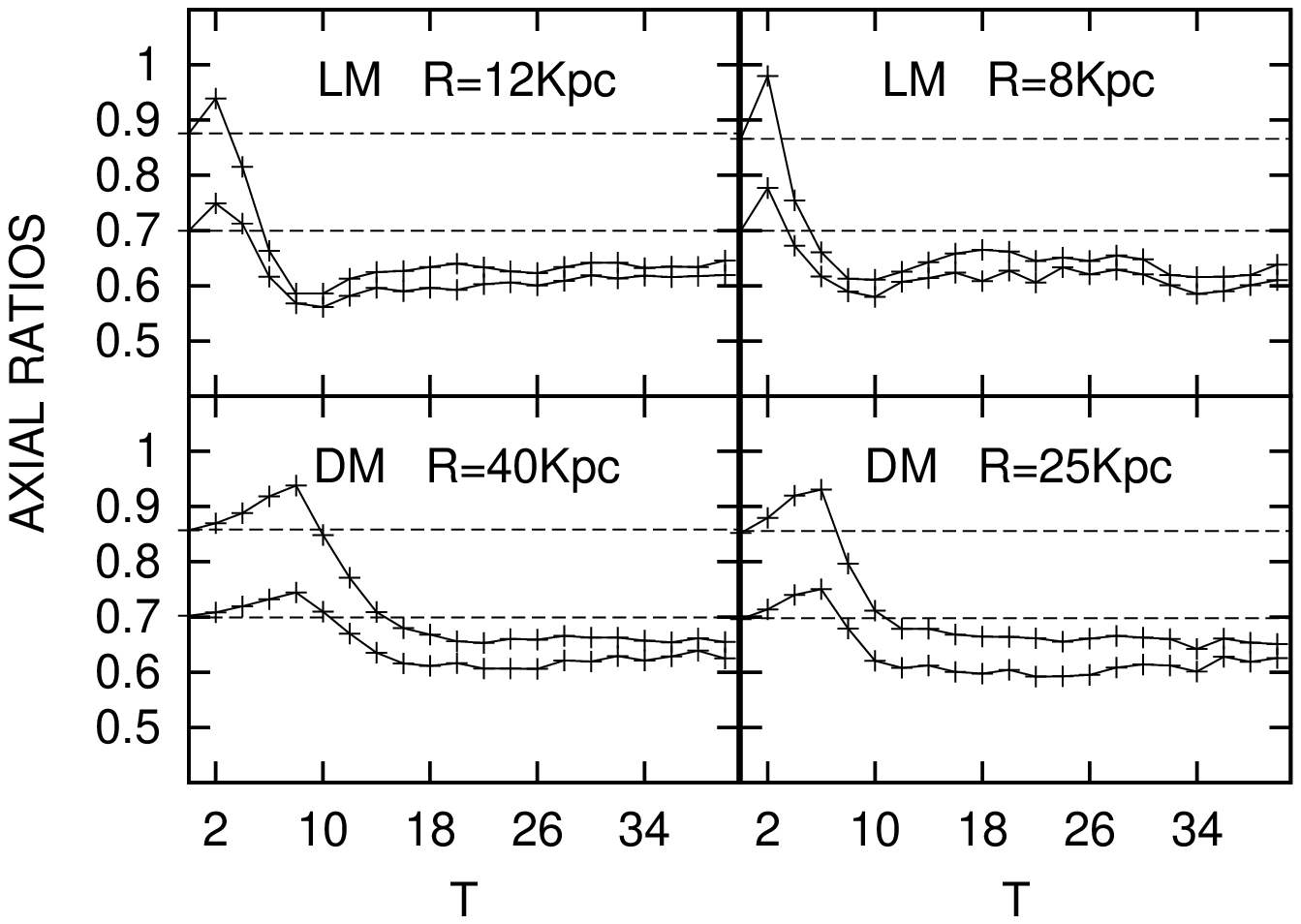} \\
                \includegraphics[angle=270,width=3.in]{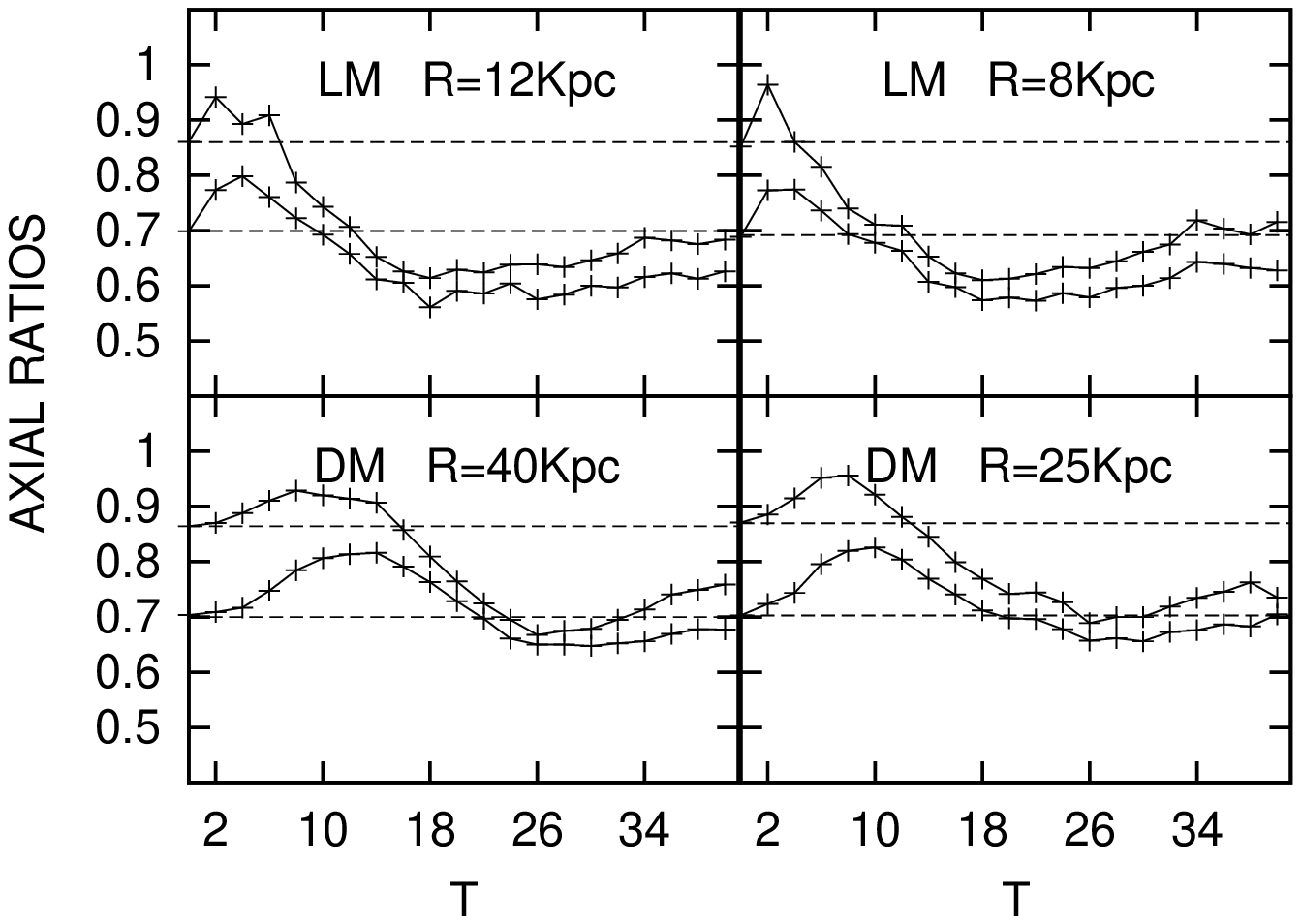} &  \includegraphics[angle=270,width=3.in]{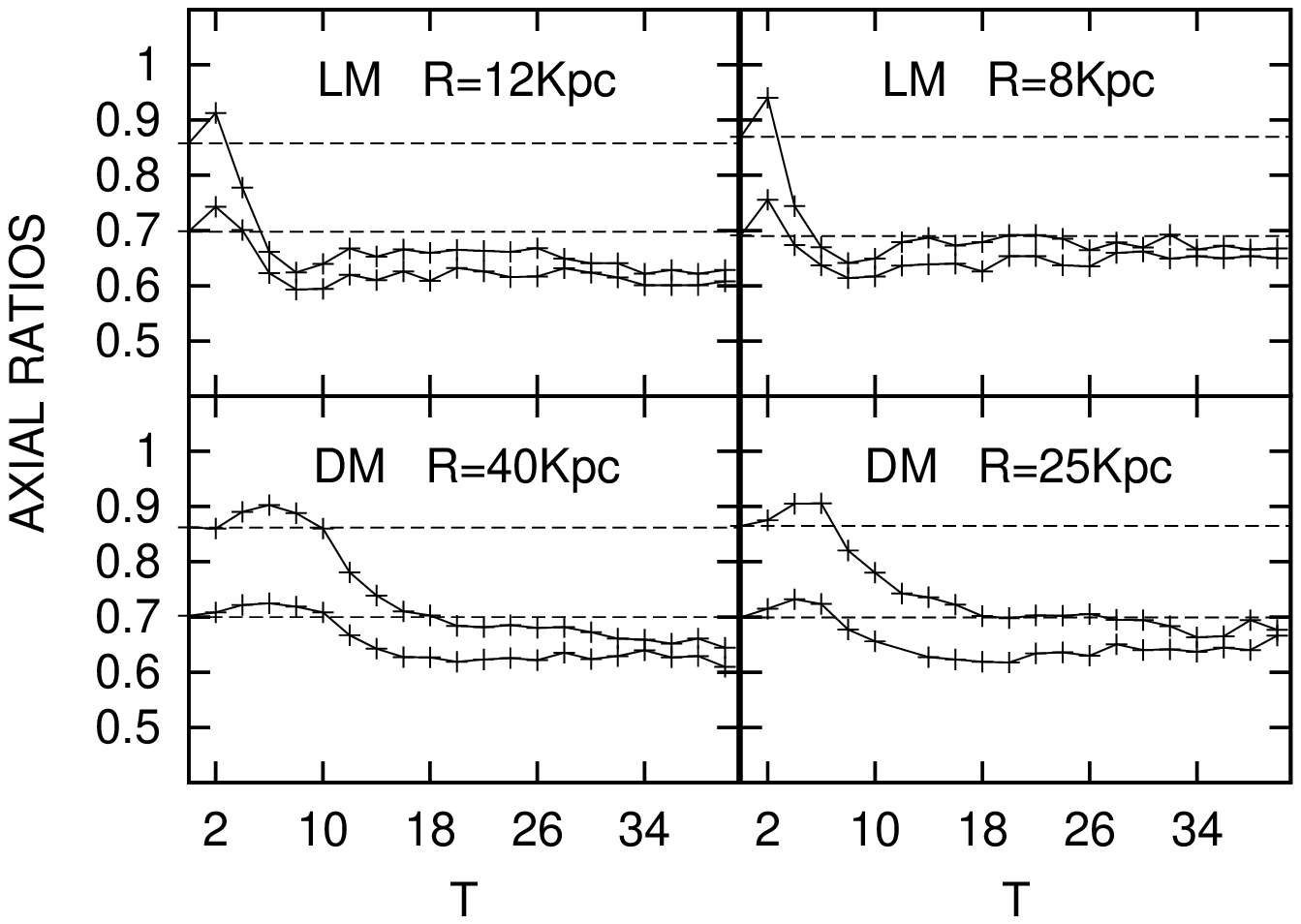}  
    \end{array}$\caption{\footnotesize{
Evolution of the axial ratios of $HL$ (top left), $LL$ (top right), 
$HL_{bis}$ (bottom left) and $LL_{bis}$ (bottom right).
R is the distance from the center where the axial ratios are evaluated. 
The times are scaled to the crossing time.
}}\label{fig6}
\end{center}
\end{figure*}

\section{$N$-Body Integrations}

\subsection{$N$-Body code}

The full $N$-body integrations were carried out using the  
{\tt TreeATD} code of \citet{Miocchi:2002}, which is a 
parallel code that uses a tree algorithm for the gravitational
force evaluation, and an individual time stepping for the leap-frog 
integrator.
{\tt TreeATD} needs three input parameters that influence the
speed and accuracy of a simulation:
the opening angle $\theta$, 
the smoothing length $\epsilon$, 
and the maximum allowed time step $\Delta t$.
We set $\epsilon=0.05$ and $\epsilon=0.1$ in the case of model $LL_{bis}$,  
$\Delta t=0.07$ and $\theta=0.7$.
These values were chosen in order to conserve energy within $0.05\%$
over the full course of the integrations.
Simulations were performed using  8 nodes of {\tt gravitySimulator},
a 32-node cluster at the Rochester Institute of Technology.

\subsection{The instability}

The $N$-body integrations revealed that both MOD1 and MOD1-bis 
represent unstable equilibria, in the sense that their axis
ratios evolve significantly.
Lagrangian radii for both models showed essentially no evolution,
indicating that the instability affects only the shapes of the models
and not the global concentration of matter.
  
Figure~\ref{fig6} illustrates the change of the axis ratios  
up to $t=711.2$ ($40$ crossing times) for the maximum-streaming models $HL$ and $HL_{bis}$.
Strong deformations appear evidently in both the luminous and 
dark components after just two crossing times.
The initial evolution is toward a more spherical shape;
the duration of  this first phase is different in the two components, 
being longer for the dark component than for the luminous component.
Final shapes are nearly prolate, with
$0.7<b/a<0.77$ and $0.65<c/a<0.7$  for the dark halo and 
$0.69<b/a<0.71$ and $0.59<c/a<0.62$ for the luminous matter.
The final contours of the projected density for model $HL$ are 
shown in Figure~\ref{fig7}.
After $\sim 25$ crossing times (dark matter) and $18$
crossing times (luminous matter), the instability 
appears to have run its course, but the system still exhibits a slow
figure rotation, as shown in Figure~\ref{fig8}.

\begin{figure*}
\begin{center}
$\begin{array}{cc}
                \includegraphics[angle=0,width=3in]{fig7a.ps}  & \includegraphics[angle=0,width=3in]{fig7b.ps} 
                    \end{array}$
\caption{Contours of the projected density after $40$ crossing times for  model $HL$.
Left panels: luminous component; right panels: dark matter component.}
\label{fig7}
\end{center}
\end{figure*}

\begin{figure*}
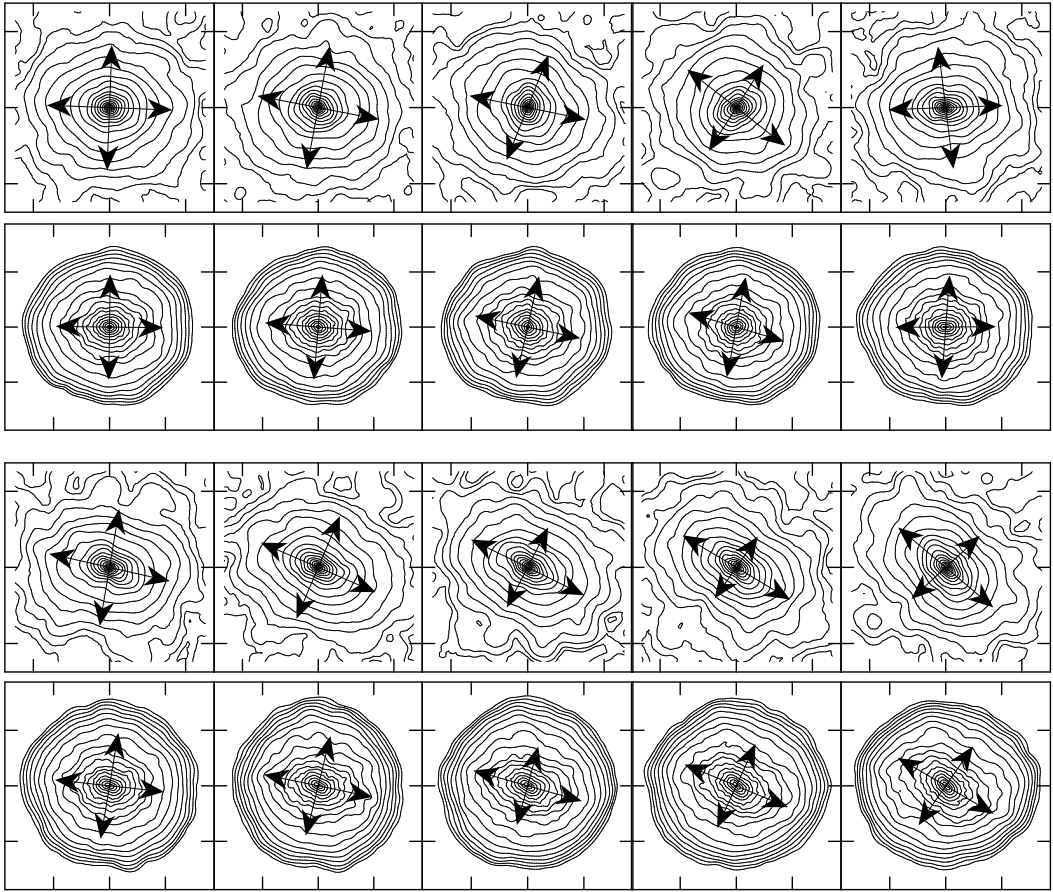

\begin{center}
$\begin{array}{c}
              \includegraphics[angle=0,width=5.5in]{fig8a.ps}  \\
              \includegraphics[angle=0,width=5.5in]{fig8b.ps} \\
\\
              \includegraphics[angle=0,width=5.5in]{fig8c.ps} \\
              \includegraphics[angle=0,width=5.5in]{fig8d.ps} \\
                    \end{array}$
\caption{Evolution of the isodensity contours in the $y-z$ plane for the 
model $HL_{bis}$; the two horizontal set of panels on the top display the configuration at 
$0,4,8,12,16t_{cross}$, then, the two bottom set of panels refer to $20,24,28,32,36t_{cross}$ 
At each time, the upper panels represent 
the luminous component while the dark matter is shown into the lower panels.
In the case of the luminous matter the linear size of each box is 
$22$ while for the dark matter it is $85$.
  The arrows represent the eigenvectors of the inertia tensor with 
the highest value of the projection on the $y-z$ plane (in most of the 
plots, the other eigenvector is approximately along the line of sight). 
 The rotation can clearly be seen after $20$ crossing times; by this time the
instability appears to have run its course.  }\label{fig8}
\end{center}
\end{figure*}

\begin{figure*}
\begin{center}
$\begin{array}{cc}
                \includegraphics[angle=0,width=3in]{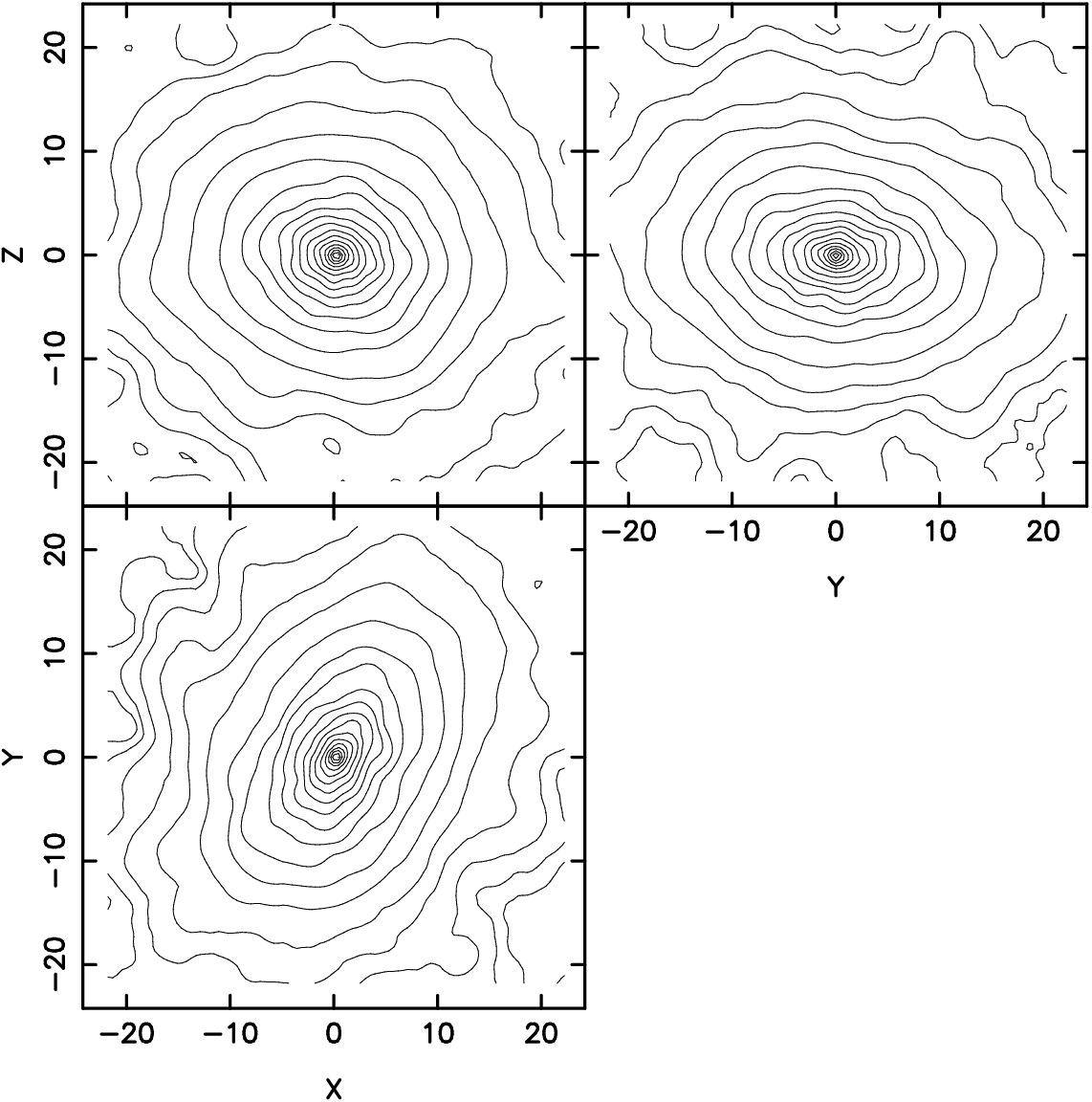}  & \includegraphics[angle=0,width=3in]{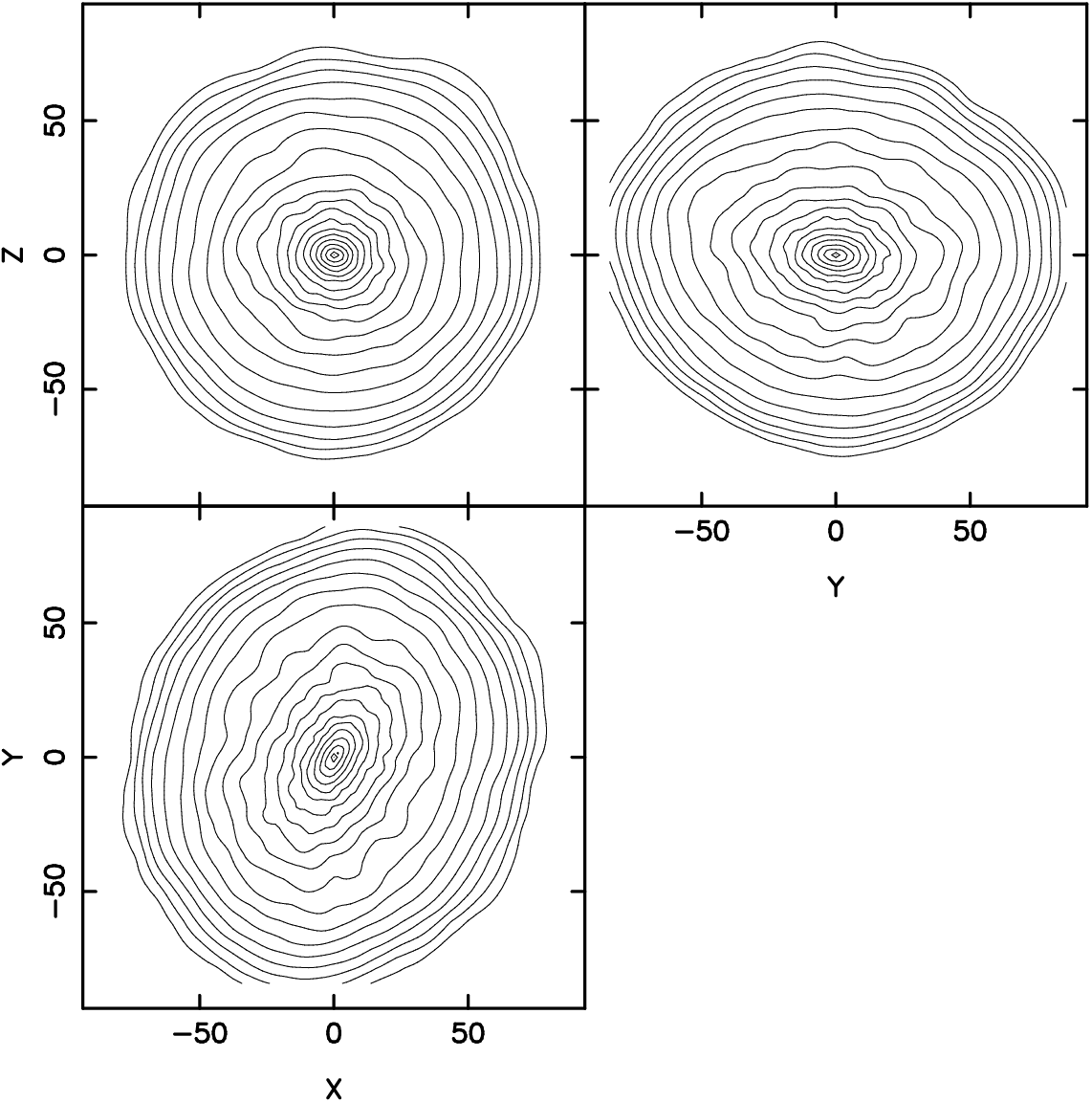} 
                    \end{array}$
\caption{Contours of the projected density at $40$ crossing times for  model $LL$.
Left panels: luminous component; right panels: dark matter component.}
\label{fig9}
\end{center}
\end{figure*}

In order to  understand better the influence of model rotation on the 
dynamical evolution, we performed a second set of simulations
for the systems $LL$ and $LL_{bis}$, which correspond to the case of 
minimum angular momentum. 
The right columns of Figure~\ref{fig6} display the evolution of the
axis ratios for models $LL$ and $LL_{bis}$, while  Figure~\ref{fig9} 
shows the contours of the projected density for both the mass 
components at the final time for model $LL$.
The results confirm that figure rotation is suppressed in these 
cases; nevertheless, dynamical evolution still leads to the 
formation of a sort of bar.
A comparison between  the evolution in the cases 
of high and low $L$ suggests that a dynamical instability occurs 
at the beginning and manifests itself completely in $\sim 20$ crossing
times.
After  this time, figure rotation is still present for $HL$ and $HL_{bis}$ 
while the $LL$ and $LL_{bis}$  systems conserve both their
shape and their orientation in space. 

Based on Figure~\ref{fig6}, the change in shape occurs sooner 
 in the case of low angular momentum;
for models $LL$ and $LL_{bis}$, the ``stable'' phase begins at 
$\sim 8t_{cross}$ in the luminous component  
and at $\sim 18t_{cross}$ in the dark component.
In addition, in the absence of rotation, the final elongation
in the dark matter is greater: $0.64<b/a<0.69$ and $0.6<c/a<0.67$.
Rotational motion has the  effect of breaking slightly the axisymmetry 
reached by the model.

This evolution is strongly reminiscent of the well-known
ROI seen in radially-anisotropic,
spherical models \citep{Merritt:1999}.
In spherical models, the instability causes a bar to form
with some random orientation, determined by the precise
spectrum of density inhomogeneities in the initial model.
In the triaxial case, the initial conditions are already bar-like,
but the instability chooses a new bar-like distortion to grow.
As in the spherical case, the final configuration after the
instability has run its course is close to prolate.

\section{Dependence of the Instability on the Orbital Composition}

In spherical models, the ROI is associated
with a predominance of eccentric orbits.
The instability growth rate can be reduced to zero by changing the
orbital composition toward more isotropic or tangentially-biased solutions; 
this is always possible since there are many distribution functions 
$f(E,J)$ that correspond to the same density profile $\rho(r)$.
Likewise, in the triaxial geometry, the Schwarzschild method can yield
a variety of orbital solutions consistent with a specified 
mass model, and these different solutions will generally have
different stability properties.
Here we show that the instability described above in the triaxial
models can indeed be effectively 
suppressed by reducing the number of semi-radial (box) orbits in 
the self-consistent solutions.
This result reinforces our hypothesis that the instability is intrinsically
similar in character to the ROI, and is also of 
physical significance for real galaxies, as discussed in \S7.

\subsection{Minimizing  the contribution from semi-radial orbits}

To  investigate the hypothesis of ROI, 
new orbital solutions  constraining the number of semi-radial 
orbits were constructed, discretized and evolved forward in time 
as $N$-body systems.
  
Following \citet{PM:2004}, the relative contributions of different
orbits to the self-consistent solutions was varied by adding a
penalty function to equations~(\ref{op_al_1}) and (\ref{op_al_2}),   
which became
\begin{eqnarray} \label{op_al_mod_1}
\chi^2_{lum}=\frac{1}{N_{cells}} \sum_{j=1}^{N_{cells}}\left(M_{j;lm}- \sum_{k=1}^{n_{orb}}C_{k;lm}B_{k,j;lm}\right)^2 +{}
\nonumber\\
 +\sum_{k=1}^{n_{orb}} C_{k;lm}W_{k;lm}~~~~~~~~~~~~~~~~~
\end{eqnarray}
and 
\begin{eqnarray} \label{op_al__mod_2}
\chi^2_{dm}=\frac{1}{N_{cells}} \sum_{j=1}^{N_{cells}}\left(M_{j;dm}- \sum_{k=1}^{n_{orb}}C_{k;dm}B_{k,j;dm}\right)^2+ {}
\nonumber\\
 +\sum_{k=1}^{n_{orb}}C_{k;dm}W_{k;dm}~~,~~~~~~~~~~~~~~~
\end{eqnarray}
respectively.
Here,  $W_{k;lm(dm)}$ is a penalty associated with the $k$th orbit 
of the luminous (dark) component;
as $W_{k;lm(dm)}$ increases, the mass contribution $C_{k;lm(dm)}$ 
of the $k$th  orbit in the model  decreases.
(We remark that the role of our penalty function is that of an 
\emph{ad hoc} numerical device and does not have any particular
physical meaning.)
We chose $W_{k;lm}= W_{k;dm}=0$ for the tube orbits and 
$W_{k;lm}\equiv W_{R;lm}>0$ and  $W_{k;dm} \equiv W_{R;dm}>0$ for the 
semi-radial orbits.
The optimization problem represented by equations (\ref{op_al_mod_1})  
and (\ref{op_al__mod_2}) was solved using the NAG routine E04NCF,
which implements an efficient method to solve solves linearly constrained linear 
least-squares problems and convex quadratic programming problems
\citep{Stoer:1971,Gill:1984}. 
The new solutions were found using the full orbital library 
corresponding to model MOD1-bis.

\begin{figure}
\begin{center}
$\begin{array}{cc}
                \includegraphics[angle=270,width=3.5in]{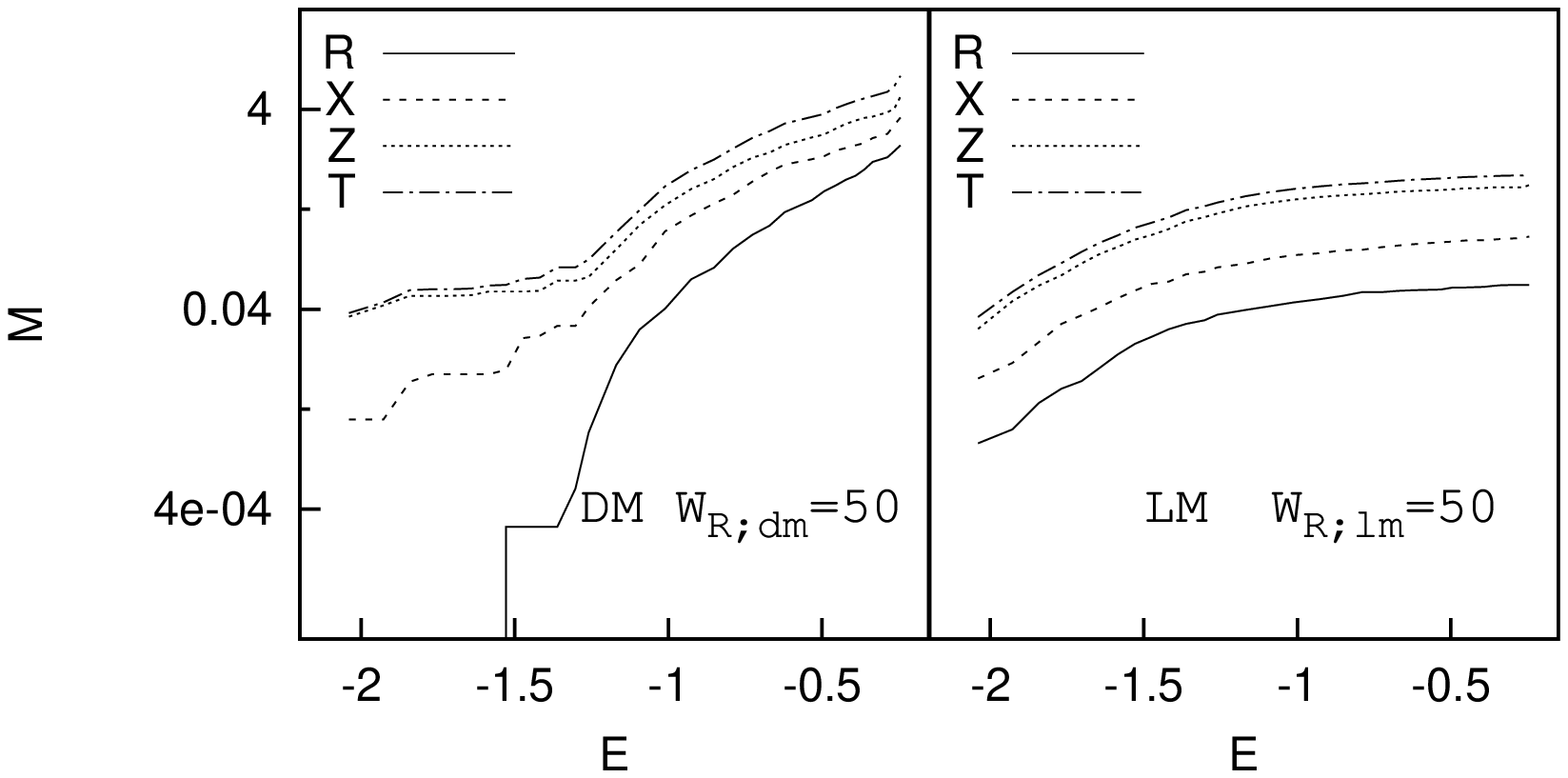} \\
                  \includegraphics[angle=270,width=3.5in]{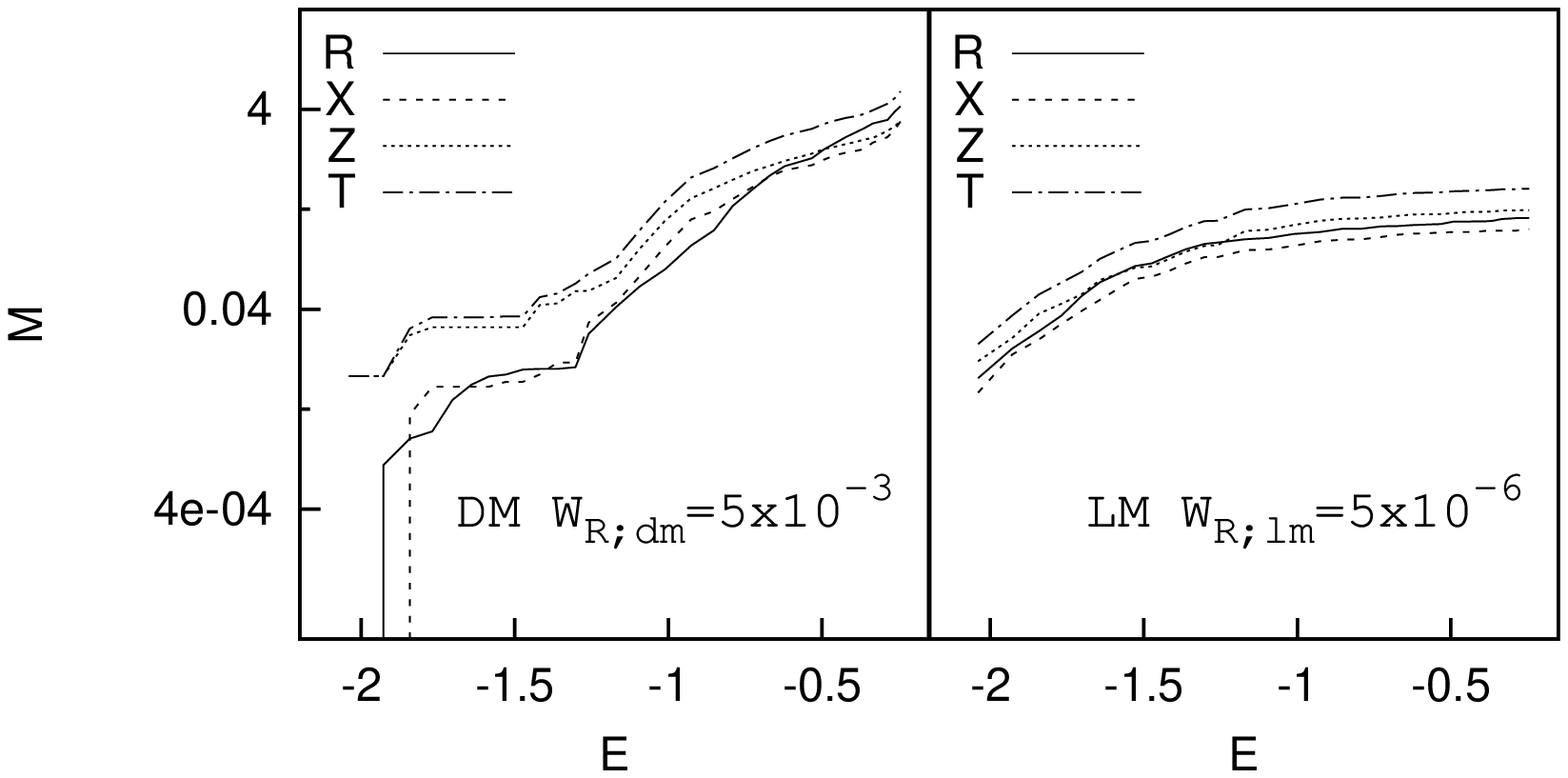} 
 \end{array}$
  \caption{  
Cumulative energy distributions (by mass) of the various orbital families 
for different values of $W_R$.
The symbols $R$, $X$, $Z$  and $T$ denote semi-radial, $X$-tube, $Z$-tube 
 and all tube orbits, respectively.}\label{fig10}
\end{center}
\end{figure}

\begin{figure}
\begin{center}
\includegraphics[angle=270,width=.5\textwidth]{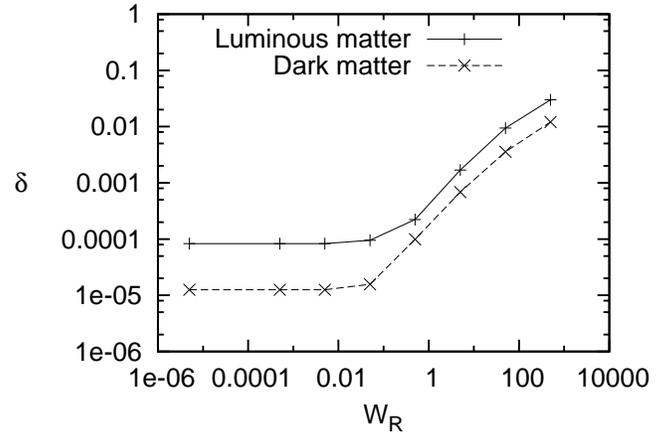}
\caption{Departure from self-consistency ($\delta$) as a function of the ``penalty''($W_R$).
}\label{fig11}
\end{center}
\end {figure}

As shown in Figure~\ref{fig10}, setting both $W_{R;dm},W_{R;lm}>0$ 
does in fact increase the number of tube orbits in the solutions
at the expense of the semi-radial orbits.
The error in the cell masses increases at increasing $W_R$, as shown 
in Figure~\ref{fig11}.
However, it is well known that \citep[e.g.][]{MF:1996}, 
the value of $\delta$ alone is not able to judge the degree 
of self-consistency of an orbital solution.
Therefore, our new solutions might still represent reasonable equilibria
even if the quality of the  fit  to the cell  masses is worse than 
that of the solutions found in CLMV07.
We return to this point below.

\subsection{Discretized models and their kinematical  properties}

We examined the properties of discretized models for four
choices of the penalty parameters:

(i)  $W_{R;dm}=W_{R;lm}=50$ (model $N1$); 

(ii) $W_{R;dm}=W_{R;lm}=5$ (model $N2$);

(iii $W_{R;dm}=5\times 10^{-3}$ and $W_{R;lm}=5\times 10^{-6}$ (model $N3$);

(iv) $W_{R;dm}=50$ and $W_{R;lm}=5\times 10^{-6}$ (model $N4$).

\noindent
$N$-body realizations were generated as explained in \S3;
the sense of circulation of the tube orbits was chosen randomly.
Each model used $2\times10^4$ luminous particles and 
$1.5\times 10^5$ dark matter particles.
Table \ref{table2} gives values of the anisotropy parameters.
As expected, $(2T_r/T_t)_{lm}$ and  $(2T_r/T_t)_{dm}$  are 
decreasing functions of $W_R$. 

\begin{figure*}
\begin{center}
$\begin{array}{cc}
                \includegraphics[angle=270,width=2.6in]{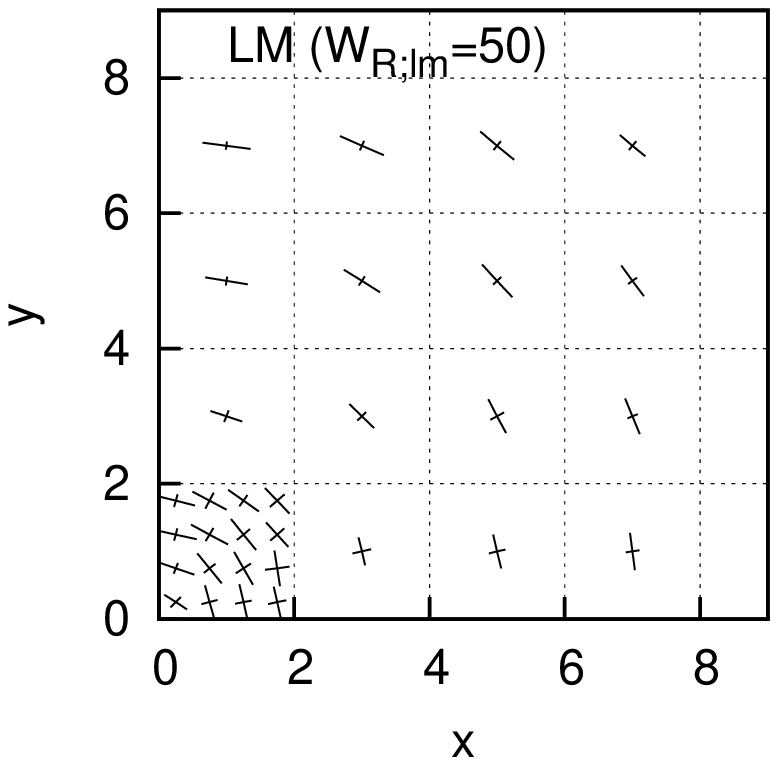} &\ \includegraphics[angle=270,width=2.6in]{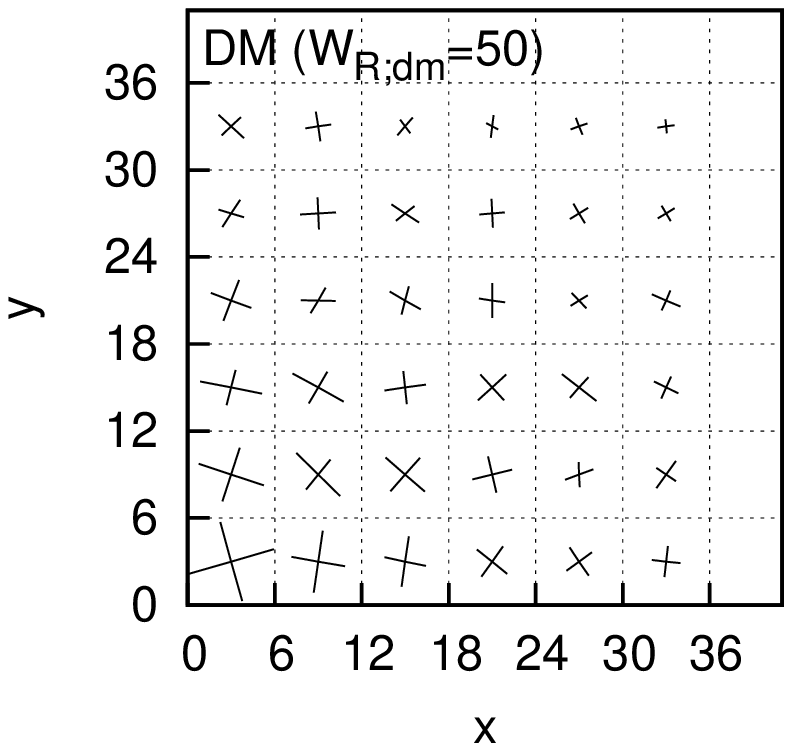}  \\
                \includegraphics[angle=270,width=2.6in]{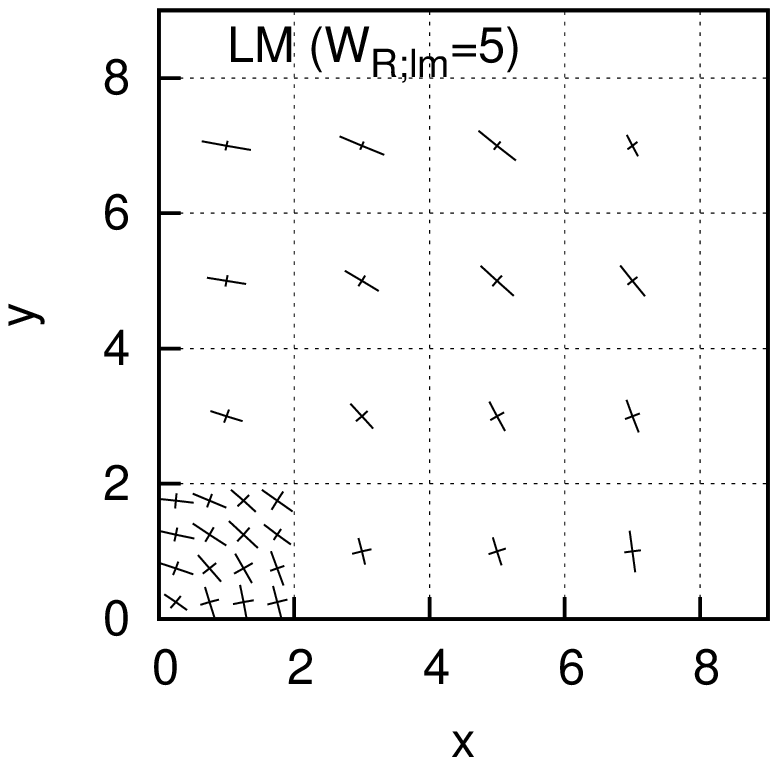} &\ \includegraphics[angle=270,width=2.6in]{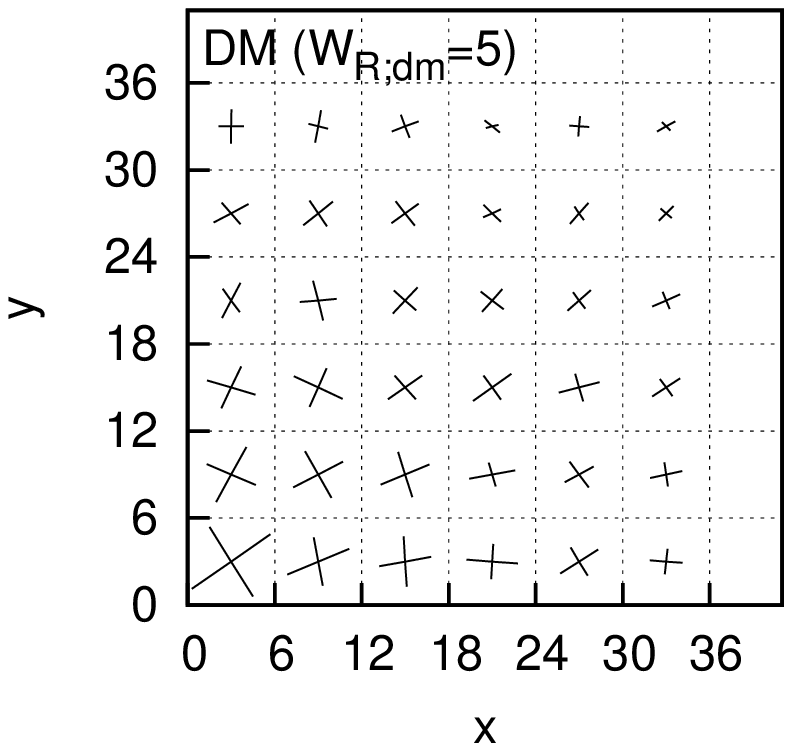}  \\
                \includegraphics[angle=270,width=2.6in]{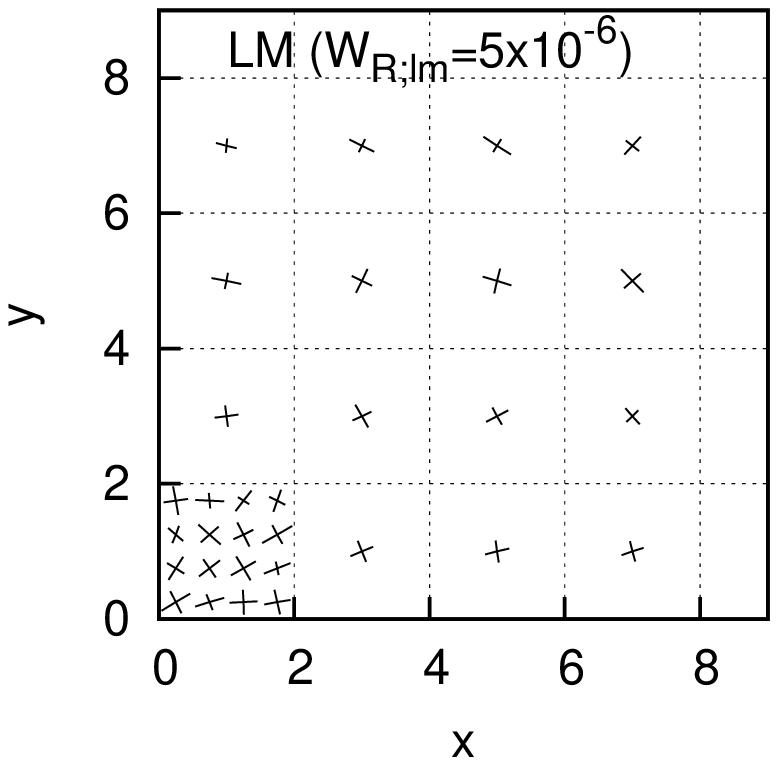} &\ \includegraphics[angle=270,width=2.6in]{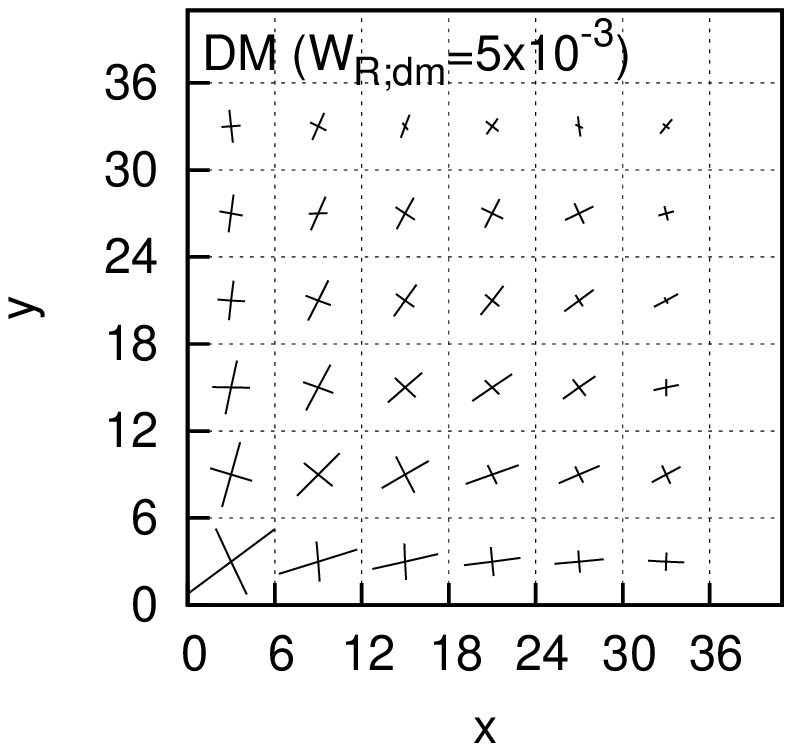}  \\               
 \end{array}$
  \caption{\footnotesize{  Velocity dispersions in model $N1$ ($W_{R,dm}=W_{R,lm}=50$), $N2$ ($W_{R,dm}=W_{R,lm}=5$) and
  $N3$ ($W_{R,dm}5\times 10^{-3},~~W_{R,lm}=5\times 10^{-6}$) . 
The length of the axes of each cross is proportional to the corresponding principal dispersion.}}\label{fig12} 
\end{center}
\end{figure*}

Figure~\ref{fig12} displays the velocity dispersions of these systems 
in the $x-y$ plane, clarifying the relation between  $W_{R}$ and the 
velocity anisotropy: 
 when $W_{R;lm}=5~~$or$~~50$, 
 the  tangential velocity dispersion is  higher than the radial
dispersion in the luminous matter. 
For $W_{R;lm}=5\times10^{-6}$, we found $\sigma_r \approx \sigma_t$.
 In the case of the dark component, when $W_{R;dm}=5~~$or$~~50$ 
the halo is nearly isotropic; 
in these systems the principal axes of the velocity ellipsoid
tend to lose their radial alignment. 
In the case $W_{R;dm}=5\times 10^{-3}$, 
the dark matter becomes  strongly anisotropic.

\begin{center}
\begin{table}
\caption{Anisotropy parameters of the new models.\label{table2}}
\begin{tabular}{lllll}
\tableline\tableline
 MODEL           &$W_{R;lm}$  & $W_{R;dm}$   &    $[2T_r/T_t]_{lm}$  & $[2T_r/T_t]_{dm}$ \\\tableline
$N1$              & $50$               & $50$               &   $0.512$    & $1.175$    \\
$N2$              & $5$                & $5$                &   $ 0.784$   & $1.335$    \\
$N3$              & $5\times 10^{-6}$  & $5\times 10^{-3}$  &   $1.220$  &   $1.754$   \\
$N4$              & $5\times 10^{-6}$  & $50$               &   $ 1.230$    & $1.174$\\
\hline
\end{tabular}
\end{table}
\end{center}

\begin{figure*}
\begin{center}
$\begin{array}{cc}
                \includegraphics[angle=270,width=3.3in]{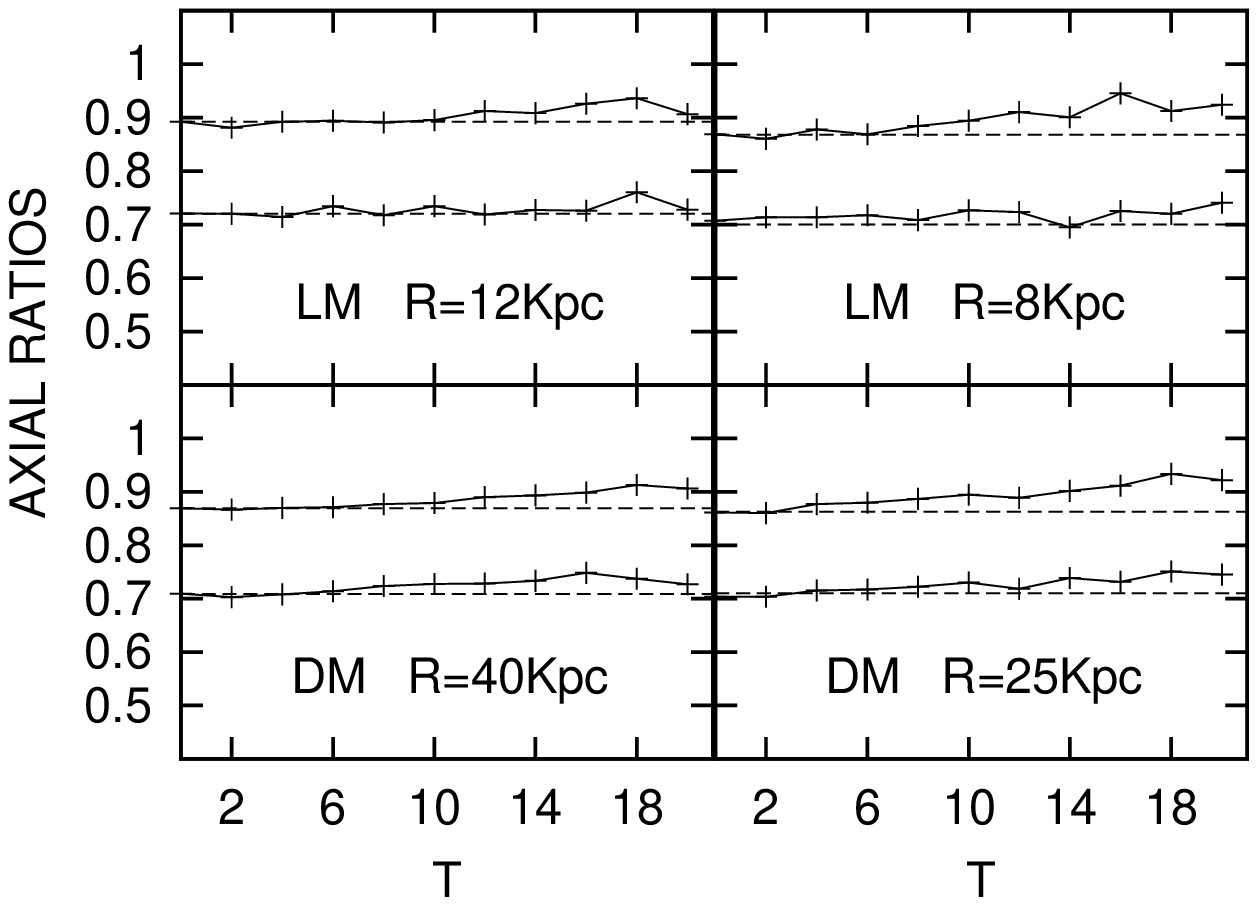}  &
                \includegraphics[angle=270,width=3.3in]{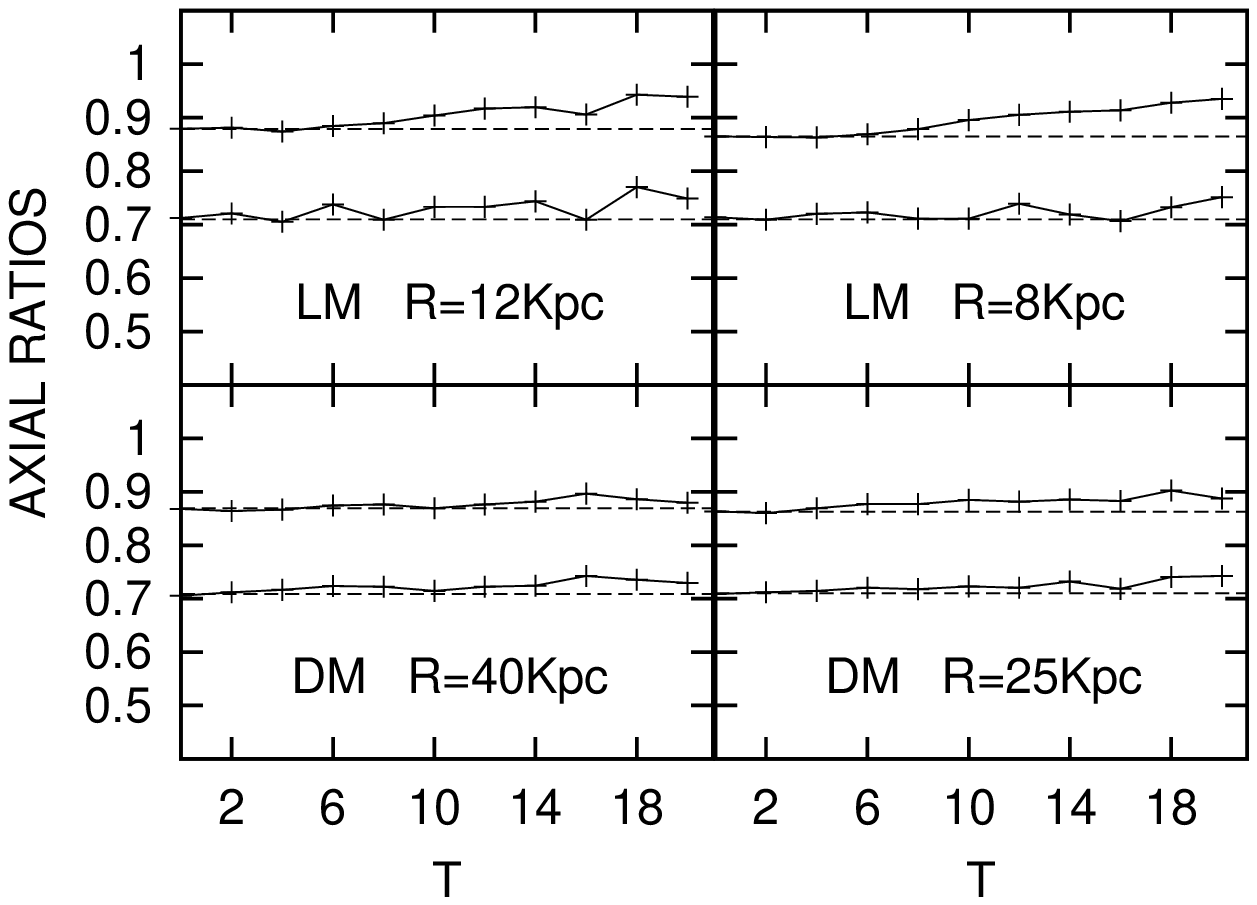}\\
                \includegraphics[angle=270,width=3.3in]{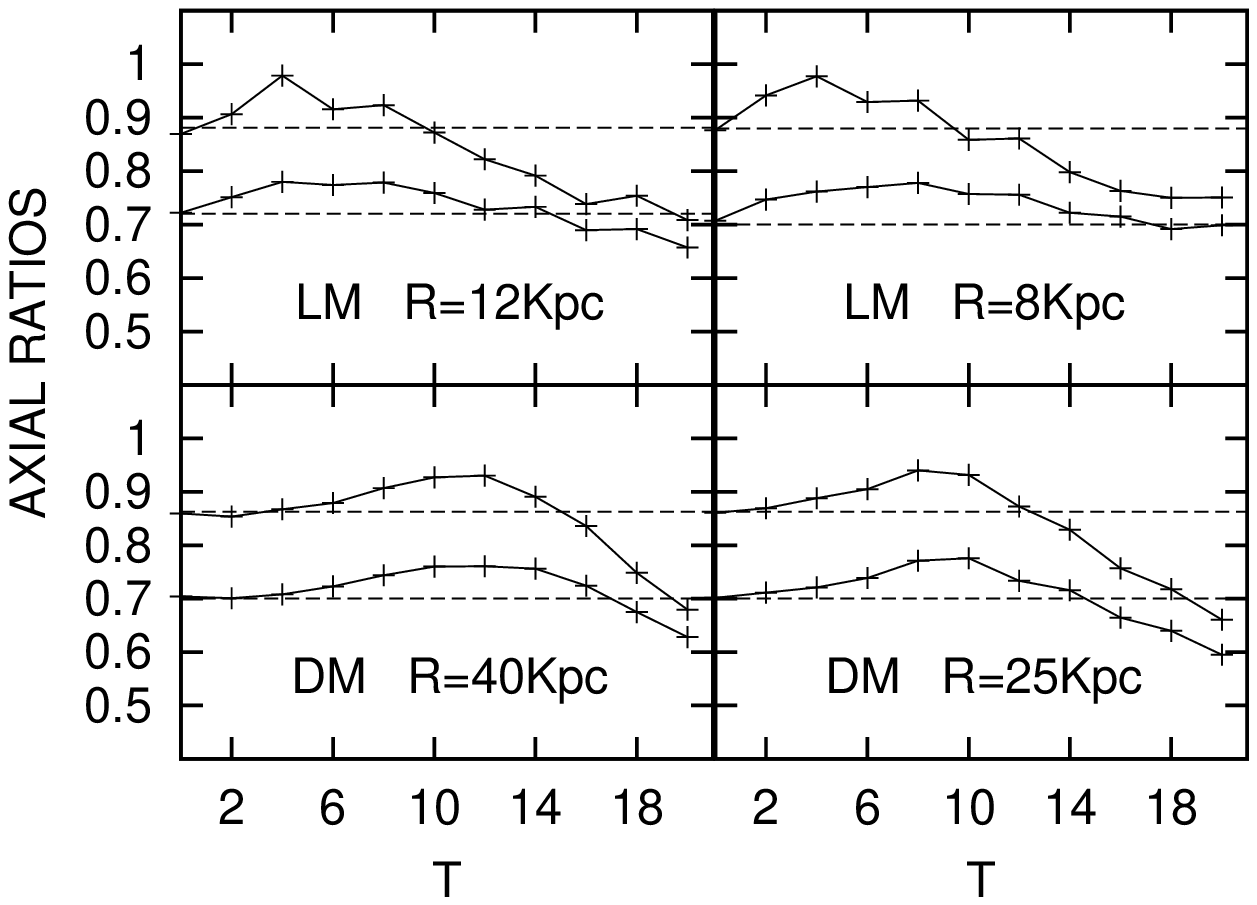}  &
                \includegraphics[angle=270,width=3.3in]{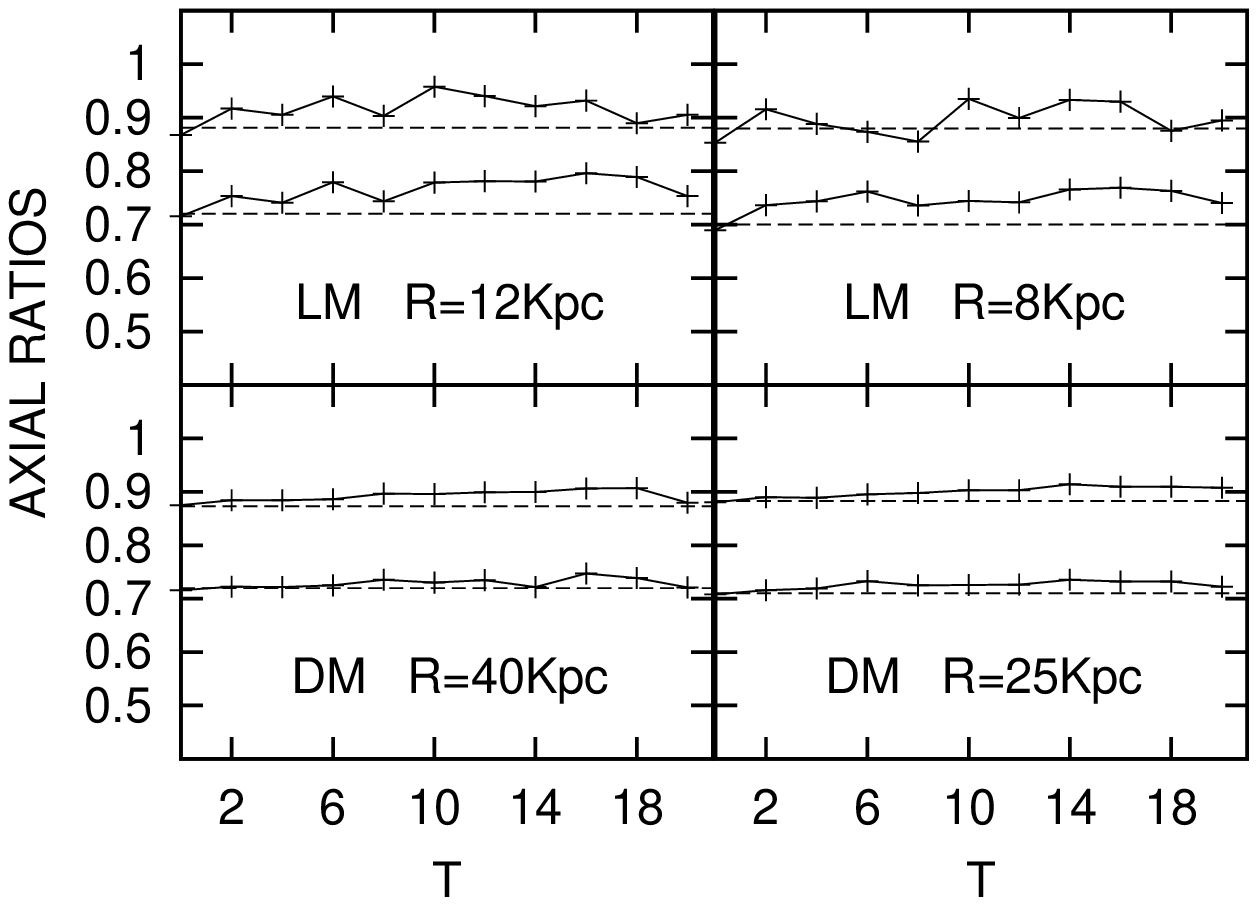} 
    \end{array}$
\caption{\footnotesize{
Evolution of the axis ratios for new models $N1$ (top left), $N2$ (top right),
$N3$ (bottom left) and $N4$ (bottom right).
R is the distance from the center where the axial ratios are evaluated. The times are scaled to the 
crossing time.
}}\label{fig13}
\end{center}
\end{figure*}

\subsection{$N$-body simulations}

\begin{figure*}
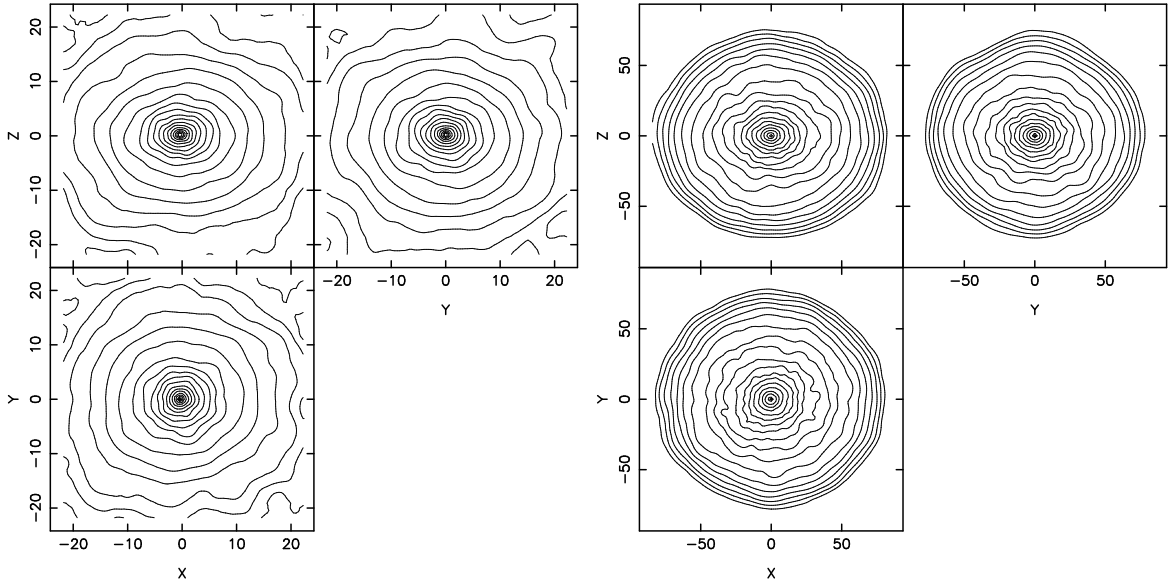

\begin{center}
$\begin{array}{cc}
                \includegraphics[angle=0,width=3in]{fig14a.ps}  & \includegraphics[angle=0,width=3in]{fig14b.ps} 
                    \end{array}$
\caption{Contours of the projected density at $20$ crossing times in model $N1$  for both the luminous matter
) and the dark matter (right).}\label{fig14}
\end{center}
\end{figure*}

\begin{figure*}
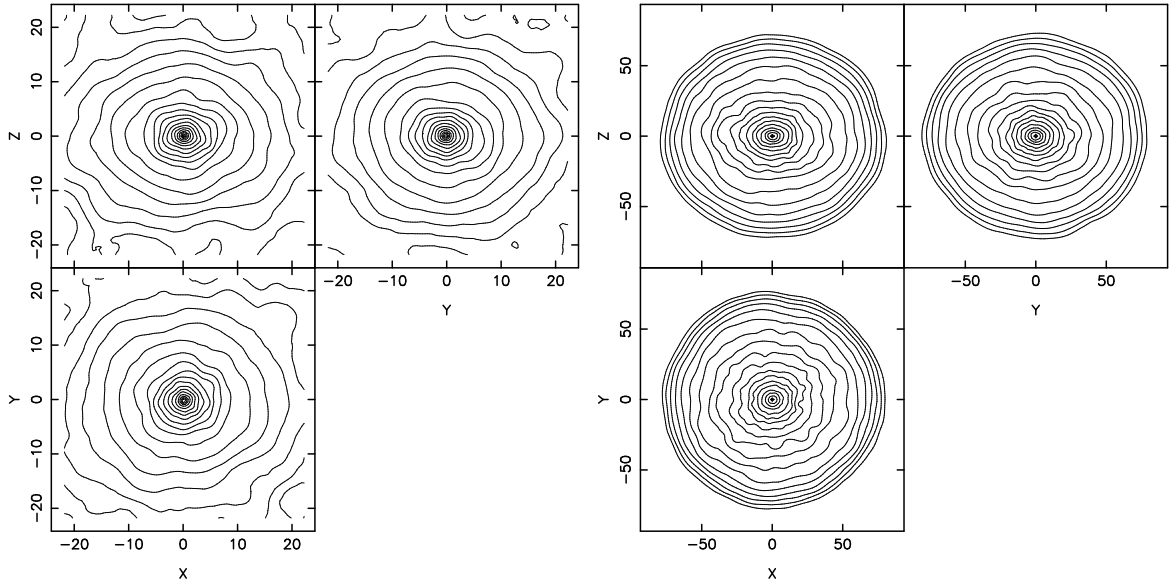

\begin{center}
$\begin{array}{cc}
                \includegraphics[angle=0,width=3in]{fig15a.ps}  & \includegraphics[angle=0,width=3in]{fig15b.ps} 
                    \end{array}$
\caption{Contours of the projected density for the luminous matter (left panels)
and the dark matter (right panels) at $20$ crossing times in model $N4$.}\label{fig15}
\end{center}
\end{figure*}

The new $N$-body systems were evolved for $20$ crossing
times, approximately the time scale of the instability to grow, 
as seen before.. 
Figure~\ref{fig13} shows the evolution of the model axis lengths.
Model $N3$ is clearly dynamically unstable, 
evolving into a prolate configuration. 
By contrast, for models $N1$, $N2$ and $N4$, the instability was
either absent or much suppressed.
Figures~\ref{fig14} and \ref{fig15} show contours of the 
projected density at $t=20t_{cross}$ for models $N1$ and $N4$ respectively; 
in these cases, the final configuration looks very similar to the 
initial one.
 
The correlation between bar formation and the value of the 
radial velocity dispersion is a clear sign that the dynamical instability
discussed in \S4.2 can be identified with the ROI.
Furthermore, the behavior of model $N4$  suggests that the instability 
disappears when  the dark halo is made isotropic, i.e.,
the instability derives mainly from anisotropy in the dark component.

We can draw the following conclusions from these integrations:

\noindent 
1. The dynamical instability disappears when the fraction of 
semi-radial orbits in the models is decreased.

\noindent
2. The stability properties of these models are determined
primarily by the kinematics of the dark matter halo. In particular,
stable configurations are obtained when  $(2T_r/T_t)_{dm} \lesssim 1.4$.

\begin{figure*}
\begin{center}
\includegraphics[angle=270,width=1\textwidth]{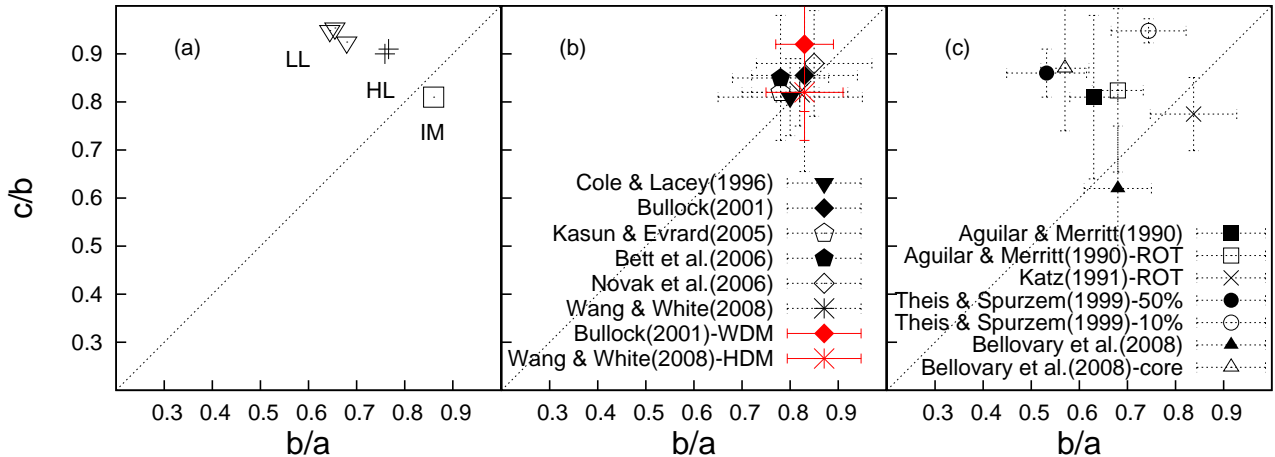}
\caption{
{\it (a)} Axis ratios of the dark matter halos of our evolved 
unstable models. 
$LL$ indicates models with low angular momentum while $HL$ refers to 
rotating models. $IM$ refers to the axial ratios of the initial model. 
{\it (b)} Mean axis ratios of dark matter halos in
simulations of structure formation in $\Lambda$CDM 
and $\Lambda$WDM cosmologies.
{\it (c)} Shapes of equilibrium $N$-body models formed by isolated
cold collapse. 
Models that were formed from rotating initial conditions are labelled
ROT.
\citet{THSP:1999} give, for a set of 9 dissipationless collapse simulations,
the axis ratios of the 50\% and 10\% of the most bound particles. 
For \citet{Bellovary:2008} we show two values of the final axis 
ratios of their initially coldest model: one refers to the innermost regions 
(core) and the other one to a larger distance, roughly the virial radius.
}\label{fig16}
\end{center}
\end {figure*}

\section{Discussion}
Large-scale simulations of structure formation have shown that 
dark matter halos, during their  evolution, develop universal 
properties, such as a characteristic density profile, 
a power-law dependence of phase space density on radius, 
a linear relation between the velocity anisotropy and the density slope 
($\beta - \gamma $ relation), and a particular distribution of
shapes and spins
\citep{DC:1991,Warren:1992,NFW:1996,Moore:1999,TN:2001,HM:2006}. 
Various authors \citep[e.g.][]{Syer:1998,Dekel:2003} have argued 
that dynamical processes during mergers may be responsible for the 
apparent universality of these relations.
On the other hand, \citet{Huss:1999} and \citet{Wang2:2008} 
have pointed out that many of
these properties are also reproduced in a 
universe where halos form via monolithic collapse, suggesting
that mergers are not essential for establishing 
the universal relations.

Some of the regularity in dark matter halo properties may be due
to dynamical instabilities, which limit the range of allowed
equilibrium states irrespective of how the halos formed.
A well-established example is the effect of bending instabilities
on the shapes of hot stellar systems: major to minor axis ratios
are limited to $\sim 3:1$ for both oblate and prolate systems 
\citep{MS:1994}.
This is a plausible explanation for the lack of elliptical
galaxies flatter than Hubble type E7 \citep{FP:1984},
and is also consistent with
the maximum elongations found for simulated dark matter halos
\citep{Bullock:2002,Allgood:2006,Bett:2007}.

The role of the ROI in establishing such ``universal'' characteristics
is less clear. 
The ROI arises naturally in halos formed via monolithic
collapse, causing otherwise spherical systems to settle into 
prolate/triaxial shapes \citep{MA:1985,AGM:1990,Huss:1999}.
The instability also reduces the dependence of the final
concentration on the initial ``temperature'' of the collapsing
cloud \citep{MA:1985,Barnes:2005,Bellovary:2008}.
But formation via mergers is qualitatively different from collapse;
and since a (spherical) model can always be rendered stable by making
its velocity distribution sufficiently isotropic,
the role that the ROI plays in determining the structure of
dark matter halos is likely to depend somewhat on the details of the halo
formation process. 

We nevertheless note that the halos formed in hierarchical cosmologies 
tend to exhibit radially-anisotropic envelopes, 
$\sigma_r/\sigma_t\approx 1.5$ \citep{COLIN:2000,Wojtak:2005,N:2008},
and that these anisotropies are similar to those of unstable models
formed via collapse, both before and after the ROI has run its course 
\citep{Bellovary:2008}, and consistent with the values that render
our two-component models unstable.
So it is plausible that the ROI or something similar is active during
the hierarchical formation of halos.

Figure~\ref{fig16} presents a weak test of this idea.
Axis ratios of our unstable halo models (panel $a$) are compared
with those of 
dark matter halos formed in various cosmological simulations 
(panel $b$) and with $N$-body models formed via simulations of isolated 
collapse (panel $c$).
As noted above, the instability has the effect of making our initially 
triaxial models more prolate and more elongated.
In the absence of rotation (LL), the final shapes are more prolate
than for typical cosmological halos.
However in the models with streaming motions (HL), 
the final axis ratios are essentially identical to the
average values found in the cosmological simulations.
A similar conclusion can be drawn from the isolated collapses
in panel (c), which also tend to be more triaxial (i.e. less
prolate) when rotation is present. 

We stress again that our unstable models could be rendered stable
by selecting different sets of orbits, in the same way that unstable
spherical models can be stabilized by making their velocity distributions
more isotropic.
The role of the ROI in structuring dark matter halos must therefore
depend somewhat on the orbital composition of halos formed in the
cosmological simulations.

\section{Conclusions}

We explored the stability properties of two, self-consistent models 
of triaxial galaxies embedded in triaxial dark matter halos.
Our results can be summarized as follows.

1. Both models were found to be dynamically unstable,
evolving toward more prolate shapes on a time scale of
$\sim 20$ crossing times.
Final shapes were approximately prolate in both components,
with short-to-long axis ratios of $\sim 0.6-0.7$. 

2. The evolution was shown not to be due to errors in construction
of the equilibrium models, nor to diffusion of chaotic orbits,
but rather to a collective mode.
On this basis we identified the instability with the ROI of spherical models.

3. Including streaming motions in the initial models leads to final
configurations that are more triaxial 
than when rotation is absent.
These final shapes are very similar to the mean shapes of
dark matter halos formed in hierarchical merger simulations.

4. When the number of box-like orbits is reduced below a certain 
threshold the dynamical instability disappears.
The presence or absence of the instability is most strongly
affected by the number of box-like orbits in 
the dark matter halo; stable configurations are obtained when 
$(2T_r/T_t)_{\rm dm}\lesssim 1.4$.
\\
\\
We thank  Linda Leccese, Paolo Miocchi and Alessandro Vicari for helpful discussions. 
This work was supported by NSF grants AST-0420920 and AST-0807910 and 
by NASA grant NNX07AH15G.

\clearpage


\begin{thebibliography}{}
\bibitem[Aguilar \& Merritt(1990)]{AGM:1990}
Aguilar, L.A., \& Merritt, D. \ 1990, ApJ, 354, 33

\bibitem[Allgood et al.(2006)]{Allgood:2006} 
Allgood, B., Flores, R.~A., Primack, J.~R., Kravtsov, A.~V., 
Wechsler, R.~H., Faltenbacher, A., \& Bullock, J.~S.\ 2006, 
\mnras, 367, 1781 

%\bibitem[Amendt \& Cuddeford(1994)]{AMCU:1994}
%Amendt, P. \& Cuddeford, P. \ 1994, \apj, 435, 93

%\bibitem[Andersen et al.(2001)]{Andersen:2001}
%Andersen, D.~R., Bershady, Matthew A., Sparke, 
%Linda S., Gallagher, J.~S., III, \& Wilcots, E.~M. \ 2001, \apj, 551, L131 

%\bibitem[Andersen \& Bershady(2002)]{Andersen:2002}
%Andersen, D., \& Bershady, M. A. \ 2002, ASPC, 275, 39

\bibitem[Antonov(1973)]{Antonov:1973}
Antonov, V.~A.\ 1973, in The Dynamics of Galaxies and Star Clusters,
ed. G. B. Omarov (Alma Ata: Nauka), 139

%\bibitem[Arnaboldi et al.(1993)]{ARNA:1993}
%Arnaboldi, M., Capaccioli, M., Cappellaro, 
%E., Held, E. V., \& Sparke, L. \ 1993, A\&A, 267, 21

\bibitem[Barnes et al.(2005)]{Barnes:2005}
Barnes, E.~I., Williams, L.~L.~R., Babul, A., \& Dalcanton, J.~J.
\ 2005, \apj, 634, 775

%\bibitem[Basilakos et al.(2000)]{BASIL:2000}
%Basilakos, S., Plionis, M., \& Maddox, S.~J. \ 2000, \mnras, 316, 779

%\bibitem[Beauvais \& Bothun(1999)]{Beauvais:1999}
%Beauvais, C. \&  Bothun, G.\ 1999, \apjl, 125, 99

\bibitem[Bellovary et al.(2008)]{Bellovary:2008}
Bellovary, J.~M., Dalcanton, J.~J., Babul, A., Quinn, T.~R., Maas, R.~W.,
Austin, C.~G., Williams, L.~L.~R., \& Barnes, E.~I.\ 2008, arXiv:0806.3434

\bibitem[Bett et al.(2007)]{Bett:2007}
Bett, P., Eke, V., Frenk, C.~S., Jenkins, A., Helly, J., \& Navarro, J.\
2007, \mnras, 376, 215

\bibitem[Bullock(2002)]{Bullock:2002} 
Bullock, J.~S.\ 2002, in The Shapes of Galaxies and Their Dark Matter Halos, 
Proceedings of the Yale Cosmology Workshop,
ed. Priyamvada Natarajan. Singapore: World Scientific, p.109	

%\bibitem[Buote \& Canizares(1996)]{BC:1996} 
%Buote, D.~A. \& Canizares, C.~R. \ 1996, \apj, 457, 565

%\bibitem[Buote \& Canizares(1998)]{BC:1998} 
%Buote, D.~A., \& Canizares, C.~R. \ 1998, \mnras, 298, 811

%\bibitem[Buote et al.(2002)]{BC:2002}	
%Buote, D.~A., Jeltema, T.~E., Canizares, C.~ R., \& Garmire, G.~P. 	
%\ 2002, \apj, 577, 183

\bibitem[Burkert(1995)]{Burkert:1995} 
Burkert, A.\ 1995, ApJ, 447, L25
	
\bibitem[Capuzzo-Dolcetta et al.(2007)]{CLMV07}
Capuzzo-Dolcetta, R., Leccese, L., Merritt, D., \& Vicari, A.\ 2007, 
\apj, 666, 165

\bibitem[Cole \& Lacey(1996)]{COLE:1996}
Cole, S., \& Lacey, C. \ 1996, \mnras, 281, 716

\bibitem[Coli\'n et al.(2000)]{COLIN:2000}
Coli\'n, P., Klypin, A. A., \& Kravtsov, A. V. \ 2000, \apj, 539, 561
 
%\bibitem[Combes \& Arnaboldi(1996)]{CA1996}
%Combes, F. \& Arnaboldi, M. \ 1996, A\&A, 305, 763

\bibitem[Dehnen(1993)]{Dehnen:93} 
Dehnen, W.\ 1993, \mnras, 265, 250

\bibitem[Dekel et al.(2003)]{Dekel:2003} 
Dekel, A., Devor, J., \& Hetzroni, G.\ 2003, \mnras, 341, 326 

\bibitem[Dubinski \& Carlberg(1991)]{DC:1991}
Dubinski, J., \& Carlberg, R.G.\ 1991, \apj, 378, 496

%\bibitem[Efthymiopoulos et al.(2006)]{Efthym:2006}
%Efthymiopoulos, C., Voglis, N., \& Kalapotharakos, C.\ 2006, arXiv:astro-ph/0610246 

\bibitem[Fehlberg(1968)]{Fehlberg:1968}
Fehlberg, E.\ 1968, NASA Tech.Rep. TR T-287

%\bibitem[Franx \& de Zeeuw(1992)]{Franx:1992}
%Franx, M., \& de Zeeuw, T.\ 1992, \apj, 392, L47

%\bibitem[Frenk et al.(1988)]{Frenk:1988}
%Frenk, C.~S., White, S.~D.~M.; Davis, M., \& Efstathiou, G.\ 1988,
%\apj, 327, 507

\bibitem[Fridman \& Polyachenko(1984)]{FP:1984} 
Fridman, A.~M., \& Polyachenko, V.~L.\ 1984, 
Physics of gravitating systems (Springer)

%\bibitem[Gentile et al.(2005)]{Gentile:2005}
%Gentile, G., Burkert, A., Salucci, P., Klein, U., 
%\& Walter, F.\ 2005, \apj, 634, L145
\bibitem[Gill(1984)]{Gill:1984}
Gill, P.E. et al.\ 1984, ACM Trans. Math. Software, 10, 282

\bibitem[Hansen \& Moore(2006)]{HM:2006} 
Hansen, S.~H., \& Moore, B.\ 2006, New Astronomy, 11, 333

%\bibitem[Hoekstra et al.(2004)]{Hoekstra:2004}
%Hoekstra, H., Yee, H.~K.~C., \& Gladders, M.~D. \
%2004, \apj, 606, 67

\bibitem[Huss et al.(1999)]{Huss:1999} 
Huss, A., Jain, B., \& Steinmetz, M.\ 1999, \apj, 517, 64 

\bibitem[Kasun \& Evrard(2005)]{Kasun:2005}	
Kasun, S.~F., \& Evrard, A.~E. \ 2005, \apj, 629, 781 


%\bibitem[Kathinka-Dalland-Evans \& Bridle(2008)]{Kathinka:2008}
%Kathinka Dalland Evans, A. \& Bridle, S.,
%\ 2008, arXiv:0806.2723

\bibitem[Katz(1991)]{Katz:1991}
Katz, N.\ 1991, \apj, 368, 325

\bibitem[Kuhlen et al.(2007)]{KDM:2007}
Kuhlen, M., Diemand, J., \& Madau, P. \ 2007, \apj, 671, 1135

%\bibitem[Kuijken \& Tremaine (1994)]{KUTR:1994}
%Kuijken, K. \&  Tremaine, S. \ 1994, \apj, 421, 178

\bibitem[Lynden-Bell(1979)]{LB:1979}
Lynden-Bell, D.\ 1979, \mnras, 187, 101

\bibitem[MacMillan et al. (2006)]{MACMI:2006}
MacMillan, J. D., Widrow, L. M., \&  Henriksen, R. \ 2006, \apj, 653, 43

%\bibitem[Massett \& Bureau(2003)]{MASSETT:2003}
%Masset, F. S. \& Bureau, M.\ 2003, \apj, 586, 152 

%\bibitem[Merrifield(2002)]{MERRF:2002}
%Merrifield, M. R. \ 2002, sgdh.conf, 170

\bibitem[Merritt(1980)]{Merritt:1980}
Merritt, D.\ 1980, \apjs, 43, 435

\bibitem[Merritt(1999)]{Merritt:1999} 
Merritt, D.\ 1999, \pasp, 111, 129 

\bibitem[Merritt \& Aguilar(1985)]{MA:1985}
Merritt, D., \& Aguilar, L.\ 1985, \mnras, 217, 787 

\bibitem[Merritt \& Sellwood(1994)]{MS:1994}
Merritt, D., \& Sellwood, J.~A.\ 1994, \apj, 425, 551

\bibitem[Merritt \& Fridman(1996)]{MF:1996} 
Merritt, D., \& Fridman, T.\ 1996, \apj, 460, 136

\bibitem[Miocchi \& Capuzzo-Dolcetta(2002)]{Miocchi:2002} 
Miocchi, P., \& Capuzzo-Dolcetta R.\ 2002, A\&A, 382, 758

\bibitem[Moore et al.(1999)]{Moore:1999}	
Moore, B., Ghigna, S., Governato, F., Lake, G., Quinn, T., Stadel, 
J., \& Tozzi, P.\ 1999, \apj, 524, L19

\bibitem[Navarro et al.(1996)]{NFW:1996} 
Navarro, J.~F., Frenk, C.~S., \& White, S.~D.~M.\ 1996, \apj, 462, 563


\bibitem[Navarro et al.(2008)]{N:2008}
Navarro, J. F., Ludlow, A., Springel, V., Wang, J., Vogelsberger, M., White, S. D. M.,
Jenkins, A., Frenk, C. S., \&  Helmi, A. \ 2008, arXiv:0810.1522


\bibitem[Novak et al.(2006)]{Novak:2006}	
Novak, G.~S., Cox, T.~J., Primack, J.~R., Jonsson, P., \& Dekel, A.
\ 2006, \apj, 646, L9

%\bibitem[Olling \& Merrifield(2000)]{OM:2000}
%Olling, R.~P. \& Merrifield, M.~R. \ 2000, \mnras, 311, 3610

%\bibitem[Parker(2007)]{Parker:2007} 
%Parker, L.\ 2007, ASPC, 380, 539

%\bibitem[Piffl \& Ya(2008)]{Piffle:2008} 
%Piffl, T., \& Ya, S.~N.\ 2008, arXiv:0811.2573 

%\bibitem[Plionis et al.(1991)]{Plionis:1991}
%Plionis, M., Barrow, J.~D., \& Frenk, C.~S. \ 
%1991, \mnras, 249, 662

%\bibitem[Plionis et al.(2006)]{Plionis:2006}
%Plionis, M., Basilakos, S., \& Ragone-Figueroa, C. \
%2006, \apj, 650, 770

\bibitem[Poon \& Merritt(2004)]{PM:2004} 
Poon, M.~Y., \& Merritt, D.\ 2004, \apj, 606, 774 

\bibitem[Polyachenko \& Shukhman(1977)]{PS:1977}
Polyachenko, V.~L., \& Shukhman, I.~G.\ 1977, Soviet Astronomy Lett., 3, 134

%\bibitem[Rix(1996)]{RIX:1996}
%Rix, H.-W.\ 1996, IAUS, 169, 23

%\bibitem[Rix \& Zaritsky(1995)]{RIX:1995}
%Rix, H.~W. \& Zaritsky, D.\ 1995, \apj, 447, 82

%\bibitem[Ryden(2006)]{RYDEN:2006}
%Ryden, B. S. 2006, \apj, 641,773

%\bibitem[Sackett et al.(1994)]{SACKETT:1994}
%Sackett, P.~D., Rix, H.~W., Jarvis, B.~J., \& Freeman, K.~C.
%\ 1994, \apj, 436, 629

%\bibitem[Sackett(1999)]{SACKETT:1999}
% Sackett, P.~D. \ 1999, ASPC, 182, 393

%\bibitem[Sackett \& Pogge(1995)]{SP:1995}
%Sackett, P.~D. \& Pogge, R.~W.
%\ 1995, AIPC, 336, 141

%\bibitem[Schoenmakers et al.(1998)]{SCHOENM:1998}
%Schoenmakers, A.~., Mack, K.-H.; Lara, L.,    
%Röttgering, H.~J.~A., de Bruyn, A.~G, 
%van der Laan, H., \& Giovannini, G.\ 1998, A\&A, 336, 455

\bibitem[Schwarzschild(1979)]{Schwarzschild:1979} 
Schwarzschild, M.\ 1979, \apj, 232, 236

\bibitem[Syer \& White(1998)]{Syer:1998} 
Syer, D., \& White, S.~D.~M.\ 1998, \mnras, 293, 337

\bibitem[Smith \& Miller(1982)]{SM:1982}
Smith, B.F., \&  Miller, R.H.\ 1982, \apj, 257, 103

\bibitem[Stoer(1971)]{Stoer:1971} 
Stoer, J.\ 1971, SIAM Numer. Anal., 8, 382

\bibitem[Taylor \& Navarro(2001)]{TN:2001} 
Taylor, J.~E., \& Navarro, J.~F.\ 2001, \apj, 563, 483

\bibitem[Theis \& Spurzem(1999)]{THSP:1999}
Theis, Ch., \& Spurzem, R.\ 1999, A\&A, 341, 361

\bibitem[Toomre (1966)]{Toomre:1966}
Toomre, A.\ 1966, in Geophysical Fluid Dynamics, Notes on the 1966
Summer Study Program at the Woods Hold Oceanographic
Institution (Ref. No. 66-46) (Woods Hole: Woods Hole Oceanographic
Inst.), 111
	
%\bibitem[Van der Marel(2001)]{Marel:2001}
%Van der Marel, R. P. 2001, \aj, 122, 1827

%\bibitem[Van Hese et al.(2008)]{Hese:2008}
%Van Hese, E., Baes, M., \& Dejonghe, H.\ 2008, arXiv:0809.0901

%\bibitem[van der Marel \& van Dokkum (2007)]{Marel:2007}
%Van der Marel, R. P., \& van Dokkum, P.G. \ 2007, \apj, 597, 878

\bibitem[Wang \& White(2008)]{Wang2:2008} 
Wang, J., \& White, S.~D.~M.\ 2007, \mnras, 380, 93

%\bibitem[Wang et al.(2008)]{Wang:2008}	
%Wang, Y., Yang, X., Mo, H.~J., Li, C., van den Bosch, F.~C., Fan, Z., \& Chen, X.
%\ 2008, \mnras, 385, 1511

\bibitem[Warren et al.(1992)]{Warren:1992}	
Warren, M.~S., Quinn, P.~J., Salmon, J.~K., \& Zurek, W.~H.\ 1992,
\apj, 399, 405

\bibitem[Wojtak et al.(2005)]{Wojtak:2005}
Wojtak, R., \L okas, E.~L., Gottl\"{o}ber, S., \& 
Mamon, G. A. \ 2005, \mnras, 361, L1

%\bibitem[Weijmans et al.(2007)]{Weijmans:2007}
%Weijmans, A., Krajnović, D., Oosterloo, T.~A., Morganti, R., 
%\& de Zeeuw, P.~T.\ 2007, IAUS, 235, 147
	
%\bibitem[Whitmore et al.(1987)]{WHIT:1987}
%Whitmore, B.~C., McElroy, D.~B., \& Schweizer, F.
%\ 1987, \apj, 314, 439

%\bibitem[ZuHone et al.(2008)]{ZuHone:2008}
%ZuHone, J.~A., Lamb, D.~Q., \& Ricker, P.~M.\ 2008, arXiv:0711.0746


\end{thebibliography}
\end{document}